\documentstyle[preprint,eqsecnum,aps,prb,epsf,epsfig,rotcapt]{revtex}

\newcommand{\be}{\begin{eqnarray}}
\newcommand{\ee}{\end{eqnarray}}
\newcommand{\la}{\langle}
\newcommand{\ra}{\rangle}
\newcommand{\p}{\partial}

\newcommand{\D}{V}
\newcommand{\cl}{{\cal L}}

\newcommand{\cv}{{\cal V}}
\newcommand{\no}{\nonumber}
\newcommand{\bmath}{\begin{mathletters}}
\newcommand{\emath}{\end{mathletters}}
\newcommand{\bc}{\begin{center}}
\newcommand{\ec}{\end{center}}
\newcommand{\figdir}{.}
\newcommand{\figsize}{5 in}

\begin{document}
%\preprint{submitted to {\it J.~Chem.~Phys.}
%\hspace{4pt}}

\title{Spectral analysis of electron transfer kinetics II
%\footnote{to be submitted, \it{J. Chem. Phys.}}
}

\author{YounJoon Jung
\footnote{younjoon@mit.edu} and Jianshu Cao \footnote{jianshu@mit.edu} }
\address{Department of Chemistry, Massachusetts Institute of Technology,
Cambridge, MA 01239}

\date{\today}
\maketitle

\begin{abstract}
Electron transfer processes in Debye solvents are studied using a 
spectral analysis method recently proposed. Spectral structure of a 
nonadiabatic two-state diffusion equation is investigated to reveal various
kinetic regimes characterized by a broad range of physical parameters; 
electronic coupling, energy bias, reorganization energy, 
and solvent relaxation rate. 
Within this unified framework, 
several kinetic behaviors of the 
electron transfer kinetics, including 
adiabatic Rabi oscillation, 
crossover from the nonadiabatic to adiabatic limits, 
transition from the incoherent to coherent kinetic limits, 
and dynamic bath effect, are demonstrated and compared with 
results from previous theoretical models. 
Dynamics of the electron transfer system is also calculated with the spectral
analysis method. It is pointed out that in the large reorganization energy
case the nonadiabatic diffusion equation exhibits a non-physical behavior,
yielding a negative eigenvalue.
\end{abstract}

%\maketitle

%\newpage
%\begin{multicols}{2}
%\narrowtext
%\newpage
\section{Introduction}
\label{intro} Since the seminal work of Marcus on the nonadiabatic electron
transfer reactions, \cite{marcus-arpc-64,marcus-bba-85} a great amount of
effort has been made in the studies of the electron transfer
reaction\cite{barbara-jpc-96} with a variety of tools such as time-resolved
spectroscopies, \cite{vos-nat-93,jonas-jpc-95,arnett-jacs-95} computer
simulation methods,\cite{kuharski-jcp-88,bader-jcp-90,warshel-arpc-91} and
analytical theories.
\cite{zusman-cp-80,yakobson-cp-80,calef-jpc-83,garg-jcp-85,hynes-jpc-86,rips-jcp-87a,sparpaglione-jcp-88,yang-jcp-89,roy-jcp-94,tang-jcp-96,stockburger-jcp-96,cho-jcp-95,cao-jcp-00}
It is not surprising that there have been great interests in the studies of the
electron transfer reaction, considering that it is involved in many important
chemical and biological systems (for the most recent reviews, see
Ref.~\onlinecite{jortner-acp-99}). For example, in the photo-synthetic reaction
center, electron transfer process creates the initial charge separation which
will eventually lead to the production of the adenosine triphosphate.
\cite{gehlen-sci-94} Also, recent studies of the molecular electronics will
depend crucially on the complete understanding and controlling of the electron
transfer in the chemical systems.\cite{davis-nat-98,martini-sci-01} Another
kind of electron transfer reactions which are currently subject to extensive
studies is proton-coupled electron transfer reactions, and many theoretical
\cite{cukier-arpc-98,cukier-jpc-96,soudackov-jcp-00,shin-cp-00} and
experimental \cite{cukier-arpc-98,roberts-jacs-95} studies have been performed
on this subject.
%and Hammes-Schiffer and co-workers.\cite{cukier-arpc-98,cukier-jpc-96}
%gave the first conceptual and
%theoretical foundations of proton coupled electron transfer
%reactions based on the dielectric continuum theory,
%\cite{cukier-arpc-98,cukier-jpc-96} and recently,
%Hammes-Schiffer and co-workers made significant improvements based upon
%multi-state formulation.\cite{soudackov-jcp-00}

One of important and ubiquitous aspects of electron transfer kinetics is the
dynamic solvent effect on the electron transfer rate.
\cite{weaver-chemrev-92,chen-chemrev-98,khoshtariya-jpca-01}
%zusman-cp-80,yakobson-cp-80,calef-jpc-83,garg-jcp-85,hynes-jpc-86,rips-jcp-87a,sparpaglione-jcp-88,yang-jcp-89,roy-jcp-94,tang-jcp-96,weaver-chemrev-92,chen-chemrev-98}
As both experimental and theoretical investigations have been carried out on
electron transfer reactions in solutions, many diverse phenomena including
predictions of the original Marcus theory have been revealed, depending on
physical parameters involved in the electron transfer kinetics.
\cite{zusman-cp-80,yakobson-cp-80,calef-jpc-83,garg-jcp-85,hynes-jpc-86,rips-jcp-87a,sparpaglione-jcp-88,yang-jcp-89,roy-jcp-94,tang-jcp-96,weaver-chemrev-92,chen-chemrev-98,khoshtariya-jpca-01}

Electronic coupling constant, $V$, given by the interaction matrix element
between 
the electron donor and acceptor wavefunctions is one of the most important
physical parameters, and depending on its magnitude compared with other
parameters, various kinetic regimes are exhibited in the electron transfer
process. When the electronic coupling constant is the smallest parameter of
the electron transfer process,
the electron transfer rate is well described by
perturbation theory, which predicts the golden-rule rate,\cite{marcus-bba-85}
\be k\approx k_{\rm GR}={2\pi V^2\over\hbar}\rho_c, \label{intro1}
\ee
where $\rho_c$ is the equilibrium population of the reactant state in the
crossing regime, and this is a well known result of the {\em Marcus electron
transfer theory}. When the electronic coupling constant large enough, 
the overall reaction process is not determined by the Marcus rate, 
Eq.~(\ref{intro1}), but by the solvent diffusion rate 
describing the polarization dynamics of the solvent molecules,
\cite{zusman-cp-80}
\be k\approx k_{\rm D}\approx{\Omega}\lambda\rho_c, \ee
where $\Omega$ is the solvent relaxation rate
%in the Debye model for solvation dynamics
and $\lambda$ is the classical reorganization energy, and this case is called
the {\em solvent-controlled
limit}.\cite{frauenfelder-sci-85,wolynes-jcp-86,rips-jcp-95,rips-jcp-96}

About two decades ago, Zusman\cite{zusman-cp-80} investigated the crossover
between the Marcus and solvent-controlled regimes 
in the studies of the electron transfer reaction in Debye solvents. 
He used a mixed quantum-classical
approach where the stochastic operator is introduced to describe bath
relaxation processes occurring on two diabatic
surfaces\cite{zusman-cp-80,yakobson-cp-80,alexandrov-znatur-81} and treated the
electron transfer process as a nonadiabatic transition between them. Zusman
solved the nonadiabatic diffusion equation in the weak coupling limit $(V\ll
k_BT)$, and obtained the expression for the overall electron transfer rate,
\be k^{-1}=k_{\rm GR}^{-1}+k_{\rm D}^{-1}, \ee
which shows a transition from the Marcus to the solvent-controlled limits in the
nonadiabatic regime.

%It is known that in the solvent controlled
%regime the overall electron transfer rate is independent of the coupling
%constant due to multiple crossings in the crossing regime.
%, and there is no real adiabaticity involved in the solvent controlled regime.
%\cite{frauenfelder-sci-85,wolynes-jcp-86}

As the electronic coupling constant is increased further to be comparable to or
larger than the thermal energy, it is expected that
the electron transfer process involves an {\em adiabatic barrier crossing}.
Finally, when the electronic coupling constant is even larger so that it has
the same order of magnitude as the solvent reorganization energy, $\lambda$,
which is indeed the case for the mixed valence
compounds,\cite{lucke-jcp-97,evans-jcp-98,jung-jpca-99,golosov-jcp-01-1,golosov-jcp-01-2}
the electronic states are delocalized on the lower adiabatic surface. Due to
the delocalization nature of electronic states, an adiabatic picture is more
useful than the diabatic one for analyzing the short-time dynamics
in strongly coupled systems.\cite{cao-cpl-99} In this picture, {\em electronic
coherence} arises from Rabi oscillation between two adiabatic surfaces.

Although there have been several studies to bridge between the Marcus regime
and the solvent-controlled regime using various approaches,
\cite{zusman-cp-80,yakobson-cp-80,calef-jpc-83,garg-jcp-85,hynes-jpc-86,rips-jcp-87a,sparpaglione-jcp-88,yang-jcp-89,roy-jcp-94,tang-jcp-96,frauenfelder-sci-85,wolynes-jcp-86,rips-jcp-95,rips-jcp-96}
few studies have discussed diverse kinetic regimes in a unified way, and often
different approaches are taken in different regimes. It is desirable to
investigate the effects of solvent dynamics on the electron transfer process in
a unified approach for various parameter regimes.

As a general approach to describing condensed phase dynamics, we recently
proposed a {\it spectral analysis method}.\cite{jung-jpca-99,cao-jcp-00} Instead of
focusing on dynamical trajectories of the reduced density matrix for
dissipative systems, this methodology investigates {\it the spectral structure of
the evolution operator for dissipative systems}, and it has been applied to the
electron transfer process in mixed valence compounds to investigate the
possibility of electronic coherence in those systems.\cite{jung-jpca-99}

In this paper we present a thorough analysis of the electron transfer kinetics
in Debye solvents based on the spectral analysis method. 
Electron transfer rate constant 
extracted from the spectral analysis 
is compared with other previous results both 
in the nonadiabatic and adiabatic regimes, and 
transition from the incoherent 
to coherent regimes is demonstrated 
by the spectral analysis method. 
When the solvent relaxation rate is very fast, 
it is found that the solvent dynamics has a
significant effect on the Marcus curve. The spectral analysis method is also
utilized as a density matrix propagation scheme. 
Preliminary results of the spectral analysis method 
focusing on symmetric reaction cases 
have been reported\cite{cao-jcp-00}. 

The rest of the paper is organized as follows. The spectral analysis method of
the nonadiabatic diffusion equation is formulated in Sec.~\ref{theo}. Two
important limiting cases are discussed in Sec.~\ref{2limit}.
In Sec.~\ref{spectral}
comprehensive analysis of the spectral structure of 
the nonadiabatic diffusion equation
is performed for a broad range of parameters, and diverse kinetic behaviors
in electron transfer reactions are identified and characterized.
We conclude in Sec.~\ref{concl} by summarizing results obtained in this work.

\section{Theory}
\label{theo} We consider two electronic states, $|1\ra$ and $|2\ra$,
which represent the electron
donor and acceptor sites of the electron transfer system, respectively, and they are
coupled to each other via the electronic coupling matrix element, $V$.
Moreover, each electronic state is coupled to bath degrees of freedom.

There have been extensive studies of the solvent effect on electron transfer
dynamics in literature with various approaches.
\cite{zusman-cp-80,yakobson-cp-80,calef-jpc-83,garg-jcp-85,hynes-jpc-86,rips-jcp-87a,sparpaglione-jcp-88,yang-jcp-89,roy-jcp-94,tang-jcp-96,frauenfelder-sci-85,wolynes-jcp-86,rips-jcp-95,rips-jcp-96}
%Formally, one can consider the full density matrix which describes the dynamics
%of the system and solvent molecules on equal footing. However, it is clearly
%impossible to consider all the bath degrees of freedom explicitly, therefore we
%need to average out the bath degrees of freedom to get a reduced description
%such that we can only consider the dynamics of the system with the effects of
%bath included.
Considering that electron transfer processes are usually probed at room
temperature in polar solvents, one can treat the bath degrees of freedom
classically. Zusman and Yakobson-Burshtein
\cite{zusman-cp-80,yakobson-cp-80} proposed a mixed quantum-classical evolution
equation of the reduced density matrix, $\rho(E,t)$, independently,
to investigate the solvent effect on electron transfer,
%the Then, the spin-boson Hamiltonian can be invoked to derive a two-level
%classical equation of motion,\cite{garg-jcp-85} which was first proposed by
%Zusman and by Yakobson and Burshtein, \cite{zusman-cp-80,yakobson-cp-80}
%
\be {\p\over\p t}\rho(E,t)=\cl\rho(E,t)=(\cl_B+i\cv)\rho(E,t). \label{zus} \ee
%
%Here, the diagonal and off-diagonal matrix elements of the reduced density matrix $\rho(E,t)$
%represent the populations of the electronic states and the coherence between them, respectively,
Here, $E$ is the solvent polarization energy which plays a role of the
reaction coordinate as first noticed by Marcus.\cite{marcus-arpc-64}
%when the system is found at the state with bath polarization energy between $E$ and
%$E+dE$ at time $t$.
$\cal{L}$ and $\cal{V}$ represent operators for the solvent relaxation dynamics
and for the electronic transition between two states, respectively.
%
%matrix elements are given by \be \cl_B=\left (\begin{array}{cccc}
%         \cl_{11} & 0 & 0 & 0 \\
%              0 & \cl_{22} & 0 & 0 \\
%             0 & 0 & \cl_{12} & 0 \\
%             0 & 0 & 0 & \cl_{21}
%             \end{array} \right ),  \label{Lr}
%ee
%nd
%\be
%\cv={1\over\hbar}\left (\begin{array}{cccc}
%        0 & 0 & V & -V \\
%            0 & 0 & -V & V \\
%%            V & -V & -\w_{12} & 0 \\
%            -V & V & 0 & \w_{12}
%           \end{array} \right ). \label{V}
%\ee
Explicitly, Eq.~(\ref{zus}) is written in terms of the density matrix elements,
%\bmath
\be
\dot\rho_{11} &=& \cl_{11} \rho_{11} + i V (\rho_{12} - \rho_{21}), \label{2a} \\
\dot\rho_{22} &=& \cl_{22} \rho_{22} - i V (\rho_{12} - \rho_{21}), \label{2b} \\
\dot\rho_{12} &=& \cl_{12} \rho_{12} - i  \omega_{12} \rho_{12} +
i {V\over \hbar} (\rho_{11}-\rho_{22}), \label{2c} \\
\dot\rho_{21} &=& \cl_{21} \rho_{21} + i  \omega_{12} \rho_{21} -
i {V\over \hbar} (\rho_{11}-\rho_{22}), \label{2d}
\ee
%\emath
Here, the diagonal and off-diagonal matrix elements of the reduced density matrix $\rho(E,t)$
represent populations of the electronic states and coherences between them, respectively,
and $\cl_{ij}$'s describe the relaxation process of classical bath over the free
energy surfaces,
with  $\cl_{ii}$ defined on the free energy surface for the $i$th
electronic state, and $\cl_{12}$ and $\cl_{21}$ defined on the
averaged free energy surface.
The functional form for the free energy surface in the electron transfer system 
is usually harmonic,\cite{kuharski-jcp-88,onuchic-jcp-93}
\be
U_1(E)&=& {(E+\lambda)^2 \over {4 \lambda}},  \label{5a} \\
U_2(E)&=& {(E-\lambda)^2 \over {4 \lambda}} + \epsilon, \label{5b}
\ee
where $\lambda$ is the reorganization energy,
%which is related to the parameters in Eq.~(\ref{sb}),
%\be
%\lambda=\sum_{\alpha}{{c_{\alpha}^2}\over{2m_{\alpha}\omega_{\alpha}^2}}
%       ={1\over\pi}\int d\omega{J(\omega)\over\omega}, \label{lambda}
%\ee
and we assume $\epsilon<0$ without loss of generality.
It is convenient to define
${\overline U}$ and $\hbar\omega_{12}$ are the average  and the difference
of the two free energy surfaces, respectively,
\be
{\overline U}(E)={{U_1(E)+U_2(E)}\over{2}}={E^2+\lambda^2\over 4\lambda}+{\epsilon\over 2}, \label{4a} \\
\hbar\omega_{12}(E)=U_1(E)-U_2(E)=E-\epsilon. \label{4b}
\ee
We set $\hbar=1$ for simplicity henceforth.
This set of a mixed
quantum-classical two-state equation has been previously derived by several
authors\cite{garg-jcp-85,sparpaglione-jcp-88,yang-jcp-89} starting from the
spin-boson
Hamiltonian.\cite{leggett-rmp-87,song-jcp-93,topaler-jcp-94,wang-jcp-99,weiss-quant-92}

We note that many chemically and biologically important electron transfer
processes take place in an overdamped solvent environment.
Then, the bath relaxation operators in
Eqs.~(\ref{2a})-(\ref{2d}) are modeled by one-dimensional Fokker-Planck operators
$\cl_{ij}$,
\be &&\cl_{ii} = D_E {\p \over \p E} \left ({\p \over \p E} + \beta
{\p U_i(E)
\over \p E} \right ), \label{3a} \\
&&\cl_{12} = \cl_{21} = {{\cl_{11}+\cl_{22}} \over 2}=D_E {\p \over \p E}
\left ({\p \over \p E} + \beta {\p {\overline U}(E) \over \p E} \right), \label{3b}
\ee
where $\beta={1/(k_BT)}$.
%${\overline U}$ and $\hbar\omega_{12}$ are the average  and the difference
%of the two free energy surfaces, respectively,
%\be
%{\overline U}(E)={{U_1(E)+U_2(E)}\over{2}}={E^2+\lambda^2\over 4\lambda}+{\epsilon\over 2}, \label{4a} \\
%\omega_{12}(E)=U_1(E)-U_2(E)=E-\epsilon. \label{4b}
%\ee

The Fokker-Planck equation models the relaxation process of the
density matrix element as a diffusion process in the energy space and various
parameters are identified; energy diffusion constant,
$D_{\rm E}=\Omega\Delta^2$, 
fluctuation of the solvent polarization energy, $\Delta^2={\la E^2\ra}=2\lambda
k_{\rm B}T$, and characteristic timescale of a Debye solvent $\tau_{\rm D}=1/\Omega$.
%
%\be D_E=\Omega\Delta^2,
%\label{7}
%\ee
%
%where  $\Delta^2$ is the mean square fluctuation of the solvent
%polarization energy, \be \Delta^2={\la E^2\ra}=2\lambda k_BT, \label{6} \ee
%and
%$\tau_D=1/\Omega$ is the characteristic timescale of the Debye solvent.
The correlation function of the solvent polarization energy is given by a 
single exponential form in a Debye solvent, 
\be C(t)=\la E(t)E(0)\ra=\Delta^2 \exp(-\Omega t). \ee
Note that since the nuclear dynamics is modeled by 
the Fokker-Planck operator, 
the possibility of the vibrational coherence is not considered  in this model of
electron transfer dynamics.
%It is  worthwhile to mention that one can obtain
%the nonadiabatic diffusion equation starting from the spin-boson Hamiltonian,
%by first deriving the evolution equation for quantum dissipative dynamics,
% and then taking the semi-classical limit using the Wigner distribution
%functions, and finally assuming the overdamped diffusion limit.\cite{garg-jcp-85,yang-jcp-89}

We investigate the spectral structure of the nonadiabatic diffusion operator, $\cl$, by
calculating eigenvalues, $\{Z_{\nu}\}$, and corresponding right and left 
eigenfunctions, $\{|\psi_{\nu}^R\ra\}$ and $\{\la \psi_{\nu}^{L}|\}$, 
%from the eigenvalue equation, 
%The right and left eigenfunctions corresponding to the 
%eigenvalue $Z_{\nu}$ are obtained, 
%
\be \cl |\psi_{\nu}^R \ra &=& -Z_{\nu}|\psi_{\nu}^R
\ra,
\label{8a} \\
\la \psi_{\nu}^L|  \cl &=& -Z_{\nu}\la \psi_{\nu}^L |. \label{8b} \ee
Because the nonadiabatic Liouville operator is non-Hermitian,
the eigenvalues are generally given by complex values, and the right and left
eigenfunctions corresponding to the same eigenvalue are not simply the
Hermitian conjugate to each other.\cite{simons-cp-73}

The method of eigenfunction solution is well known for the diffusion process on
the harmonic potential energy surface as discussed in App.~\ref{app-single}.
\cite{risken-fpe-84} 
Unlike the diffusion problem on the single potential energy surface, however,
there have been limited studies on the nonadiabatic diffusion problem involving
more than a single potential energy surface. Cukier and co-workers
have calculated the electron transfer rate by calculating the lowest eigenvalue
of the nonadiabatic diffusion equation in the weak-coupling regime.\cite{yang-jcp-89}
%An important issue in solving
%the nonadiabatic diffusion equation for electron transfer is the choice of the
%basis functions.  
%since  three different free energy surfaces are involved in
%Eqs.~(\ref{2a})-(\ref{2d}): two diabatic surfaces for the population density
%matrix elements and one average surface for the coherence density matrix
%element. 

In this paper, the eigenfunctions of $\cl_{12}$ are used as our basis
set to represent the nonadiabatic diffusion equation. In principle, one could
have chosen the eigenfunctions of $\cl_{11}$ or $\cl_{22}$ as basis functions;
however, in that case one has to evaluate appropriate Franck-Condon factors
when calculating the coupling matrix elements. The Fokker-Planck operator
$\cl_{12}$ is defined on the averaged single harmonic potential centered at $E=0$,
and its eigenvalue solutions are obtained following a similar procedure 
in App.~\ref{app-single},
\be
\cl_{12}| \phi^R_n \ra&=&-n \Omega | \phi^R_n \ra, \label{13a} \\
\la \phi^L_n|\cl_{12} &=&-n \Omega \la \phi^L_n |, \label{13b} \ee
where $n=0,1,2,\cdots$, and the $n$th right and left eigenfunctions for $\cl_{12}$ are given by
\be \phi^R_n(E)&=&{\exp\left(-{{E^2}\over{2 \Delta^2}}\right)\over{(2^n
n!)^{1\over2}(2\pi\Delta^2)^{1\over4}}}
H_n\left({E\over {\sqrt{2} \Delta}} \right), \label{14} \\
\phi^L_n(E)&=&{1\over{(2^n n!)^{1\over2}(2\pi\Delta^2)^{1\over4}}}
H_n\left({E\over {\sqrt{2} \Delta}}\right), \label{15} \ee
with $H_n$ being the
$n$th order Hermite polynomial. 
%As shown below, this choice of the basis set is
%convenient for our purpose.

We separate real and imaginary parts of the coherence density matrix, namely,
$u=$Re$\rho_{12}$ and $v=$Im$\rho_{12}$, and rewrite Eqs.~(\ref{2a})-(\ref{2d}) as
%\bmath
\be
\dot\rho_{11} &=& (\cl_{12}+\delta\cl) \rho_{11} - 2 V v, \label{9a} \\
\dot\rho_{22} &=& (\cl_{12}-\delta\cl) \rho_{22} + 2 V v, \label{9b} \\
\dot u &=& \cl_{12} u + \omega_{12} v, \label{9c} \\
\dot v &=& \cl_{12} v - \omega_{12} u  + V (\rho_{11}-\rho_{22}), \label{9d}
\ee
%\emath
where 
$\delta\cl\ = (\cl_{11}-\cl_{22})/2$.
%. \label{10}
%\ee
Then, all the relevant operators in Eqs.~(\ref{9a})-(\ref{9d}) can be evaluated
in terms of the basis functions, 
%right and left eigenfunctions of $\cl_{12}$,
\be
\la \phi_{n}^{L} | \cl_{12}  | \phi_{m}^{R} \ra&=&-n\Omega\delta_{nm}, \label{16a}  \\
\la \phi_{n}^{L} | \delta\cl | \phi_{m}^{R} \ra&=&-\Omega\sqrt{{\lambda}
\over{2k_BT}}\sqrt{m+1}\delta_{n,m+1}, \label{16b}  \\
\la  \phi_{n}^{L} | \omega_{12} | \phi_{m}^{R} \ra&=&\sqrt{2\lambda k_BT}
(\sqrt{m}\delta_{n,m-1}+\sqrt{m+1} \delta_{n,m+1})
-\epsilon\delta_{nm}, \label{16c} \\
\la  \phi_{n}^{L} | V| \phi_{m}^{R} \ra&=&V\delta_{nm}. \label{16d} \ee 
%where we used the Condon approximation.
With this basis set, we can expand the density matrix elements as
%\bmath
\be
\rho_{11}(E,t)&=&\sum_{n=0}^{\infty} a_{n}(t)\phi_{n}^{R}(E), \label{17a} \\
\rho_{22}(E,t)&=&\sum_{n=0}^{\infty} b_{n}(t)\phi_{n}^{R}(E), \label{17b} \\
u(E,t)&=&\sum_{n=0}^{\infty} c_{n}(t)\phi_{n}^{R}(E),  \label{17c} \\
v(E,t)&=&\sum_{n=0}^{\infty} d_{n}(t)\phi_{n}^{R}(E).  \label{17d}
\ee
%\emath
Substituting Eqs.~(\ref{17a})-(\ref{17d})
into the right eigenvalue equation Eq.~(\ref{8a}),
we have the following coupled eigenvalue equations,
%\bmath
\be
-Z_{\nu} a_{n}&=&-n\Omega a_{n}-\Omega\sqrt{\lambda\over{2k_BT}}\sqrt{n}a_{n-1}
-2 V d_{n}, \label{18a} \\
-Z_{\nu} b_{n}&=&-n\Omega b_{n}+\Omega\sqrt{\lambda\over{2k_BT}}\sqrt{n}b_{n-1}
+2 V d_{n}, \label{18b} \\
-Z_{\nu} c_{n}&=&-n\Omega c_{n}+\sqrt{2\lambda k_BT}(\sqrt{n+1}d_{n+1}+\sqrt{n}d_{n-1})
-\epsilon d_{n}, \label{18c} \\
-Z_{\nu} d_{n}&=&-n\Omega d_{n}-\sqrt{2\lambda k_BT}(\sqrt{n+1}c_{n+1}+\sqrt{n}c_{n-1})
+\epsilon c_{n} +V (a_{n}-b_{n}), \label{18d}
\ee
%\emath
which is an explicit basis set representation for the two-state diffusion
operator in Eqs.~(\ref{2a})-(\ref{2d}). 
Eigenvalue equations for the left eigenvector 
in Eq.~(\ref{8b}) can be written by making the transpose of
Eqs.~(\ref{18a})-(\ref{18d}). 
Diagonalizing the $4N\times 4N$ matrix ($N$ =
number of basis functions) defined in Eqs.~(\ref{18a})-(\ref{18d}), we obtain
the eigenvalues $Z_{\nu}$ and the corresponding eigenvectors of the
nonadiabatic diffusion operator, \be
|\psi_{\nu}^{R}\ra&=&\sum_{n=0}^{\infty} R_{n\nu}|\phi_{n}^{R}\ra, \label{eig1} \\
\la\psi_{\nu}^{L}|&=&\sum_{n=0}^{\infty} L_{\nu n}\la \phi_{n}^{L}|, \label{eig2}
\ee
where $R_{n\nu}$ and $L_{\nu n}$ are elements of the transformation matrices.

In general, due to  the non-Hermitian nature of the nonadiabatic diffusion
operator, the left and right eigenfunctions {\em do not form an orthogonal set
by themselves}. However, when the eigenvalues are all nondegenerate, the left
and right eigenfunctions form an orthogonal and complete set in {\em a dual
Hilbert space}.\cite{hatano-prb-98,dahmen-condmat-99,dahmen-jmb-00} Explicitly,
we have
\be \sum_{n=0}^{\infty}L_{\nu n}R_{n\nu^{'}}&=&\delta_{\nu\nu^{'}}, \label{19} \\
 \sum_{\nu=0}^{\infty} R_{n\nu}L_{\nu m}&=&\delta_{nm}, \label{20} \ee
for the orthogonality and the completeness relation, respectively. Using these properties, we can
construct the real time propagator for the operator $\cl$ as
\be G(t)=\sum_{\nu=0}^{\infty}|\psi_{\nu}^{R}\ra \la\psi_{\nu}^{L}|
e^{-Z_{\nu}t}, \label{21} \ee
and express the time evolution of the density matrix as an eigenfunction
expansion,
\be |\rho(t)\ra=G(t)|\rho(0)\ra=\sum_{\nu=0}^{\infty}|\psi_{\nu}^{R}\ra
\la\psi_{\nu}^{L}|\rho(0)\ra e^{-Z_{\nu}t}. \label{22} \ee
Note that the right and left eigenfunctions play asymmetric roles in the
construction of the propagator for the non-Hermitian operator.
%Hence, the
%eigenvalue solution  to the two-state nonadiabatic diffusion equation leads to  a
%complete description of  electron transfer dynamics.

\section{Limiting Cases}
\label{2limit}

In order to compare the eigenvalue solution developed in this work with
previous theoretical predictions available in different kinetic regimes, we
briefly discuss two limiting cases which have been studied extensively in the
literature.\cite{zusman-cp-80,yakobson-cp-80,calef-jpc-83,garg-jcp-85,hynes-jpc-86,rips-jcp-87a,sparpaglione-jcp-88,yang-jcp-89,roy-jcp-94,tang-jcp-96}
Instead of giving detailed derivations, we will briefly mention the solutions
in these limiting cases, relegating details to the App.~\ref{app-nonad}.

\subsection{Nonadiabatic Regime : Weak Coupling Case}
\label{nalimit} When the electronic coupling matrix element $V$ is very small,
we can reduce 
the full nonadiabatic diffusion equation, Eqs.~(\ref{2a})-(\ref{2d}),
into the population evolution equation, and detailed derivations can be found
in Refs.~\onlinecite{sparpaglione-jcp-88,yang-jcp-89} (see also App.~\ref{app-nonad}).
%
%\be
%{{\p{\vec{\rho}}}\over\p t}=({\bf L}-{\bf K}){\vec{\rho}} \ee where
%$\vec{\rho}=(\rho_{11},\rho_{22})^T$ and \be {\bf L}&=&\left (\begin{array}{cc}
%          \cl_{11} & 0 \\
%          0 & \cl_{22}
%          \end{array} \right), \\
%{\bf K}&=&\left (\begin{array}{cc}
%          K & -K \\
%          -K & K
%          \end{array} \right).
%\ee
%
Kinetic equations for the population elements of $\rho$ can be approximately
written as
\be
{{\p\rho_{11}}\over\p t}&=&\cl_{11}\rho_{11}-K(\rho_{11}-\rho_{22}),  \\
{{\p\rho_{22}}\over\p t}&=&\cl_{22}\rho_{22}-K(\rho_{22}-\rho_{11}), 
\label{red1} \ee
where $K(E)$ is the rate kernel given in Eq.~(\ref{ke}). 
%in terms of the coherent Green function
%$G(E,t|E_0)$ defined in Eq.~(\ref{g12}),\cite{sparpaglione-jcp-88,yang-jcp-89}
%%
%\be K(E)=2V^2{\rm Re}\int_0^{\infty}{\rm d}\tau\int_{-\infty}^{\infty}{\rm
%d}E_0 G_{12}(E,\tau|E_0). \label{ke} \ee
%
%We proceed to solve this coupled equations by introducing the Green's function
%matrix ${\bf G}(E,t|E_0)$ and ${\bf G_0}(E,t|E_0)$, which propagate the
%population density with and without ${\bf K}$, respectively, \be
%\hat{\bf G}(E,s|E_0)&=&(s-{\bf L}-{\bf K})^{-1}\delta(E-E_0), \\
%\hat{\bf G}_{0}(E,s|E_0)&=&(s-{\bf L})^{-1}\delta(E-E_0).
%\ee
A dynamical quantity usually measured in the electron transfer kinetics
experiment is the total population, $P_{i}(t)$,
in each electronic state rather than the polarization
energy dependent population, $\rho_{ii}(E,t)$,
\be P_{i}(t)=\int_{-\infty}^{\infty} {\rm d} E{\rho}_{i}(E,t). \ee
Using the projection operator method and making
a time-scale separation approximation, it can be shown that the kinetic process
between two electronic states is described by the time-independent rate constant instead
of the rate kernel in the weak coupling limit,
\cite{sparpaglione-jcp-88,yang-jcp-89}(see also App.~\ref{app-nonad}). Then the kinetic
equation for $P_{i}$ is given by\cite{sparpaglione-jcp-88,yang-jcp-89}
\be
\left (\begin{array}{c} \dot P_1(t) \\
           \dot P_2(t) \end{array} \right)=
         \left (\begin{array}{cc}
          -k^{1} & k^{2} \\
          k^{1}  & -k^{2}
          \end{array} \right) \left (\begin{array}{c} P_1(t) \\ P_2(t)  \end{array} \right).
\ee
Here $k^{1}$ and $k^{2}$ are forward and backward rate constants and they
are related to the nonadiabatic transition rates $k_{\rm NA}^{i}$ and solvent diffusion rates 
$k_{\rm D}^{i}$ by\cite{sparpaglione-jcp-88,yang-jcp-89}
\be k^{i}&=&{k_{\rm NA}^{i}\over 1 +k_{\rm NA}^{1}/k_{\rm D}^{1}+k_{\rm
NA}^{2}/k_{\rm D}^{2}}, \label{ki}
\\
k_{\rm NA}&=&k_{\rm NA}^{1}+k_{\rm NA}^{2}, \label{kNAtot}
\\
k_{\rm ND}&=&k^{1}+k^{2}={k_{\rm NA}\over{1+k_{\rm NA}^{1}/k_{\rm D}^{1}+k_{\rm
NA}^{2}/k_{\rm D}^{2}}}, \label{ktot}
\ee
Recently, Eq.~(\ref{ktot}) has also been obtained for the symmetric reaction case in the
weak coupling limit by calculating the first excited eigenvalue explicitly via
the Goldstone theorem.\cite{cao-jcp-00}

Eq.~(\ref{ktot}) has a form of an overall relaxation rate for consecutive
reactions, and it involves two different types of rate processes; nonadiabatic
transition and solvent diffusion rate. 
Nonadiabatic transition rate constants $k_{\rm NA}^{1}$
and $k_{\rm NA 2}^{2}$ are forward and backward quantum transition rates
between the two electronic states and are calculated in terms of the coherent
Green's functions, ${G}_{12}(E,t|E_0)$ and
${G}_{21}(E,t|E_0)$\cite{sparpaglione-jcp-88,yang-jcp-89}(see also App.~\ref{app-nonad}),
\be 
k_{\rm NA}^{i}=2\pi V^2{\rm Re}\int_0^\infty {\rm d} t
e^{i(\epsilon+\lambda)t-g(t)}, \label{kNA} \ee
%
%where the equilibrium population distributions at each electronic state are
%given by the Boltzmann distributions,
%%
%\be \rho_{ii}^{\rm eq}(E)&=&{1\over\sqrt{2\pi\Delta^2}}\exp\left[-{(E\pm
%\lambda)^2\over 2\Delta^2}\right], \label{eq1} \ee
%%
%with $+$ for $i=1$ and $-$ for $i=2$,
where the bath correlation function $g(t)$ is given by
\be g(t)=\left[\left(\Delta\over\Omega\right)^2i\pm i {\lambda\over\Omega}\right]
\left[\exp(-\Omega t)+\Omega t-1\right]. \label{g}
\ee
with $+(-)$ sign for $i=1(2)$.
%
%$k_{\rm NA}^{2}$ is obtained from Eqs.~(\ref{kf}) and (\ref{g}) by the
%substitution of $\lambda\to -\lambda$.
When the bath dynamics is slow such that 
% compared with the magnitude of the thermal 
%fluctuation of the polarization energy, 
%
%\be 
$\Omega \ll \Delta$, 
%\ee 
Eq.~(\ref{kNA}) reduces to the standard Marcus result(see App.~\ref{app-nonad}),  
\be k_{\rm NA}^{i}&\approx& k_{\rm GR}^{i}, 
=2\pi V^2 \rho_{i}^{\rm eq}(\epsilon), \label{kmar} \\ 
%=V^2\sqrt{\pi\over{\lambda k_BT}}
%\exp\left[-{(\epsilon\pm\lambda)^2\over{4\lambda k_BT}}\right], \label{kmar} \\
k_{\rm NA}&\approx& k_{\rm GR}^{1}+k_{\rm GR}^{2}. \label{kmartot}
\ee
%\\
%k_{\rm NA}&\approx& k_{\rm GR}^{1}+k_{\rm GR}^{2} \label{k12mar} \ee
%
where $\rho_{i}^{\rm eq}(\epsilon)=
e^{-(\lambda\pm\epsilon)^2/2\Delta^2}/{\sqrt{2\pi \Delta^2}}$ 
is the equilibrium population 
of the $i$th state at the crossing point $E=\epsilon$. 
When $\Omega \sim \Delta$, 
the nonadiabatic transition rate shows a deviation 
from the Marcus rate, which will be discussed later in details. 

Solvent diffusion rate constants, $k_{\rm D}^{1}$ and $k_{\rm D}^{2}$,
describe solvent relaxation processes that equilibrate
the nonequilibrium wavepacket created at the
crossing region, $E_c=\epsilon$, during the electron transfer process. 
They can be written in terms of the population
Green's functions ${G}_{11}(E,t|E_0)$ and
${G}_{22}(E,t|E_0)$,\cite{hynes-jpc-86,sparpaglione-jcp-88,yang-jcp-89}
\be (k_{\rm D}^{i})^{-1}&=&{1\over\rho_{ii}^{\rm
eq}(\epsilon)}\int_0^{\infty}{\rm
d}t[G_{ii}(\epsilon,t|\epsilon)-G_{ii}(\epsilon,\infty|\epsilon)]. \label{kd1}
%(k_{D}^2)^{-1}&=&{1\over\rho_{22}^{eq}(\epsilon)}\int_0^{\infty}dt[G_{22}(\epsilon,t|\epsilon)-G_{22}(\epsilon,\infty|\epsilon)].    \label{kd2}
\ee
Noting that $G_{ii}(\epsilon,\infty|\epsilon)=\rho^{\rm eq}_{i}(\epsilon)$
is given by the equilibrium
population distribution of the $i$th state at $E=\epsilon$, 
the solvent diffusion rates are identified as the inverse of the mean survival time of the relative nonequilibrium
population created at the crossing point, $E_c=\epsilon$.

The boundary between the Marcus regime and the solvent-controlled regime in 
the nonadiabatic limit 
is estimated when comparing the Marcus rate in 
Eq.~(\ref{kmar}),  
and the solvent-diffusion rate in Eq.~(\ref{kd11}), which leads to
\be
2\pi V^2 \sim |\epsilon\pm\lambda| \Omega. 
\ee
The nonadiabatic regime is established when 
the timescale of the off-diagonal density matrix 
is much faster than that of the diagonal ones, yielding the following condition,
\cite{zusman-cp-80,cao-jcp-00} 
\be
V \ll D_{\rm E}^{1/3}={(2\lambda k_B T \Omega)}^{1/3}. 
\ee
in addition to the weak coupling condition, $V\ll k_B T$, 
for the validity of the nonadiabatic regime. 

\subsection{Adiabatic Regime : Strong Coupling Case}
\label{adlimit} When the electronic coupling constant is comparable to or larger than 
the thermal energy $\beta V\gtrsim 1$,
it is more adequate to describe electron transfer in 
an adiabatic representation. Given a specific bath configuration
characterized by the value of $E$, the electronic part of the Hamiltonian
in the diabatic representation is
%the free energy and electronic coupling for
%the donor and acceptor states is
given by
\be H_{\rm d}(E)=\left(\begin{array}{cc}
      U_1(E) & V \\
        V & U_2(E) \end{array}\right). \label{hp}
\ee
After the diagonalization of $H_{\rm d}(E)$ for a given bath configuration $E$, we
can obtain two adiabatic surfaces,
\be U_{\pm}(E)&=&{\overline U}(E)\pm{1 \over
2} \left[\omega_{12}(E)^2+4 V^2\right]^{1/2}.
%\no \\
%                &=&{E^2\over{4\lambda}}\pm{1\over 2}\left[(E-\epsilon)^2+4V^2\right]^{1/2}+{\lambda\over 4}+ {\epsilon\over 2}.
\label{upm}
\ee

The separation between two adiabatic surfaces is much larger than $k_B T$,
$U_{+}-U_{-}\ge V \gg k_B T$.
When the energy bias $\epsilon$ and the electronic coupling $V$ are not so
large compared with the reorganization energy, $\lambda$, the lower adiabatic
surface, $U_{-}$, has a well-defined double well structure.
In this case, we can consider the electron transfer process as a
diffusional barrier crossing process occurring on the lower adiabatic surface
$U_{\rm AD}=U_{-}$ described by the
following diffusion equation in the energy space,
\be {\p\rho\over \p t}=D_{\rm E}\left({\p^2\over\p E^2}+\beta{\p\over\p
E}U_{\rm AD}'\right)\rho=\cl_{\rm AD}\rho. \label{FPad}
\ee

To calculate the reaction rate for the adiabatic barrier crossing described by 
Eq.~(\ref{FPad}), one may take several different
routes. For example, one can calculate the barrier crossing rate  as the first
nonzero eigenvalue of the Fokker-Planck operator, $\cl_{\rm AD}$, 
corresponding to the lower adiabatic surface,\cite{hanggi-rmp-90}
\be \cl_{\rm AD}\psi=-Z\psi\Rightarrow k_{\rm AD}^{\rm ev}=Z_1 \label{eigad}
\ee
Another popular approach is a population-over-flux
method,\cite{hanggi-rmp-90} which obtains the barrier crossing rate by
calculating the steady-state flux at the crossing region, $E_c=\epsilon$, divided
by equilibrium population in the reactant region,
\be
 k_{\rm AD}^{\rm fp}
 %&=&D_E^{-1}\int_{-\infty}^{E_c}dE_1e^{-\beta U_{-}(E_1)}\int_{E_1}^{E_c}dE_2e^{\beta U_{-}(E_2)} \no \\
   =D_{\rm E}\left[\int_{-\infty}^{E_c}{\rm d}E_1\int_{-\infty}^{E_1}{\rm
   d}E_2e^{\beta\left[U_{-}(E_1)-U_{-}(E_2)\right]}\right]^{-1}
\label{flux}
\ee
In the symmetric reaction case ($\epsilon =0$), the adiabatic rate given in
Eq.~(\ref{flux}) can be evaluated approximately in the strong coupling limit
($\beta V \gg 1$) by expanding the integrand at $E=0$ 
and performing a Gaussian integration approximately 
to yield\cite{hynes-jpc-86,cao-jcp-00}
\be k_{\rm AD}^{\rm fp}\approx {\Omega\over \pi}\sqrt{\lambda\over 2 V}
e^{-\beta ({\lambda\over 4}-V)},  \label{kadapp} \ee
which is the Kramer's adiabatic reaction rate in the strong damping regime.\cite{hanggi-rmp-90}
%This analysis shows that the adiabatic electron transfer rate increases with
%the coupling constant due to the reduction of the barrier height.
%In the asymmetric case ($\epsilon\neq 0$),
%Eq.~(\ref{flux}) should be evaluated numerically in general.

\section{Results of Spectral Analysis}
\label{spectral}
%We present results of detailed spectral
%analysis of the nonadiabatic diffusion equation in this section.

\subsection{General Feature of Spectra}
\label{specstr}
%\subsubsection{General Feature}
Figure \ref{fig1} shows 
the eigenvalue 
spectra in three different cases of the energy bias for
various electronic couplings. Since the nonadiabatic diffusion operator is not
Hermitian, the eigenvalues are complex in general,
\be \cl\psi_{\nu}=-Z_{\nu}\psi_{\nu}=-(Z'_{\nu}+i Z''_{\nu} ) \psi_{\nu}.
\label{eigen} \ee
In order for the density matrix relax to the stationary state at long times,
the real part of the eigenvalue should satisfy $Z'_{\nu}\ge0$. Eigenvalues of
the nonadiabatic diffusion operator are classified into three different cases
with their associated kinetic behaviors in the density matrix evolution;

%\begin{array}{cccl}
 (1) $Z'_{0}=Z''_{0}=0$  :  {equilibrium}, 

 (2) $Z'_{\nu}>0, Z''_{\nu}=0$ :  {exponential decay},

 (3) $Z'_{\nu}>0, Z''_{\nu}\neq 0$  :  {damped oscillation}. \\
%\end{array}
The eigenstate with zero eigenvalue [case (1)] corresponds to the equilibrium
solution of the density matrix while those with real [case
(2)] and complex eigenvalues  [case (3)] correspond to exponential decays and
to damped oscillations in the evolution of the density matrix, respectively.
When the first nonzero real eigenvalue $Z'_{1}$ is well separated from the
other higher eigenvalues, $Z'_{1}\ll Z'_{\nu\ge 2}$, the dynamics of the
density matrix will be described by a single exponential process, and the
relaxation rate is well defined as $k=Z'_{1}$. However, if this is not the case,
the dynamics of the density matrix will generally involve multiple timescales
and coherent oscillations.

We used 800 basis functions to calculate the eigenvalues, and to remove the
finite basis set effect at higher eigenvalues, we present only lower 400
eigenvalues in Fig.~\ref{fig1}. Three different cases of the energy bias
chosen in Fig.~\ref{fig1}(a)-(c) correspond to different regimes in
nonadiabatic electron transfer theory, i.e., normal ($|\epsilon|<\lambda$ in
Fig.~\ref{fig1}(a)), activationless ($|\epsilon|=\lambda$ in
Fig.~\ref{fig1}(b)), and inverted regimes ($|\epsilon|>\lambda$ in
Fig.~\ref{fig1}(c)), respectively. As a general feature, all the spectra show
{\em tree-like structures with three major characteristic branches};
{\em one branch for the real eigenvalues in the middle, real axis, and 
two other branches for the complex conjugate eigenvalues}.
%branches in the complex planes}.
%The real eigenvalues
%appear in the middle branch without imaginary parts, and the other two
%symmetric branches lying at the left and right half planes have the complex
%conjugate paired eigenvalues due to the non-Hermitianity of the nonadiabatic
%diffusion operator.
Therefore, the spectral structure indicates that {\em multiple exponential
decays} ({\em real eigenvalues}) and {\em damped oscillations} ({\em complex
eigenvalues}) are inherent features in electron transfer processes in the
overdamped solvent. In order to infer the dynamical behavior of the density
matrix, we only need to focus on the lower part of the eigenvalue diagrams in
Fig.~1 corresponding to the eigenstates with $Z'/\Omega \lesssim 1$ since those
with $Z'/\Omega \gg 1$ will decay out very quickly.
%\begin{figure}
%\begin{center}
%\end{center}
%\vspace{1in}
%\begin{center}
%(a) \\
%\vspace{.1cm}
%\epsfxsize=2.7 in
%\rotate{\rotate{\rotate{\epsffile{\figdir/fig1a.eps}}}}
%(b) \\
%\vspace{.1cm}
%\epsfxsize=2.7 in
%\rotate{\rotate{\rotate{\epsffile{\figdir/fig1b.eps}}}}
%(c) \\
%\vspace{.1cm}
%\epsfxsize=2.7 in
%\rotate{\rotate{\rotate{\epsffile{\figdir/fig1c.eps}}}}
%\end{center}
%\caption{
%The eigenvalue diagrams are shown
%as the coupling constant is increased in the case of three different
%values of the energy gap (a)$\beta\epsilon=0$(normal regime), (b) $\beta\epsilon=10$(activationless regime),
%and (c) $\beta\epsilon=20$(inverted regime).
%The parameters are $\beta\lambda=10$ and $\beta\Omega=1$. As the coupling constant becomes large, the branching point
%in the real branch moves down, and it bifurcates horizontally by the amount of the Rabi frequency
%for the large coupling case.
%}
%\label{fig1}
%\end{figure}

(i) {\it symmetric reaction case}($\epsilon=0$) 
: In Fig.~\ref{fig1}(a), 
one can notice that the real
parts of the eigenvalues, $Z'_\nu$'s, located at the branching regime in the
eigenvalue tree decrease as the coupling constant increases up to $\beta
V\lesssim 1$, which means that the oscillatory components will persist during a
more extended period $t\sim 1/ Z'_{\nu}$ as the coupling constant increases.
When the coupling constant increases further, $\beta V\gg1$, the branching
point of the complex branches start to separates along the imaginary axis and
the separation between the two complex branches along the imaginary axis is given
by $\sim 2V$, and this 
corresponds to the Rabi oscillation frequency for the two
adiabatic states in the strong coupling regime
%, and this which will be discussed later in details.
% into which the two diabatic surfaces for the donor and
%acceptor states can be mapped. Therefore, when the coupling constant is very
%large, a suitable picture for the dynamical behavior of the system density
%matrix is the adiabatic oscillation between the donor and acceptor
%states.\cite{cao-cpl-99}

(ii) {\it asymmetric reaction cases}($\epsilon\neq 0$) : In Figs.~\ref{fig1}(b) 
($\beta |\epsilon|=10$) and (c) ($\beta |\epsilon|=20$),  
we notice that 
spectral structures show more branched behaviors in the asymmetric reaction cases 
than in the symmetric reaction case in Fig.~\ref{fig1}(a). 
%the general features in the spectra are quite similar to that of
%the symmetric case, Fig.~\ref{fig1}(a),
%although there are distinct features of the spectra in the asymmetric cases.
When comparing the zero electronic coupling cases,
$\beta V=0$, in Figs.~\ref{fig1}(a), (b), and (c),
we notice that in the asymmetric reaction cases two complex
branches separate out along the imaginary axis by the amount of the energy
bias, $\epsilon$ unlike in the symmetric reaction case.
This corresponds to the natural oscillation frequency 
of the off-diagonal density matrix elements. 
In Fig.~\ref{fig2} we have followed the evolution of the lowest 30 eigenvalues
in both the real and imaginary axes as a function of the coupling constant for
different energy bias cases. In all the three cases real parts of
the eigenvalues start as pairs of doubly degenerate states 
separated from each other by the solvent relaxation rate, $\beta\Omega=1$, when $V=0$. 
As the coupling constant increases, the degeneracies of the real eigenvalues
are first lifted at small coupling constants, and bifurcation and
coalescence of eigenvalues occur at large coupling constants. The coalescence
corresponds to the case where two closely separated real eigenvalues become
complex conjugate with the same $Z'$, and this indicates the transition 
from the incoherent, rate process to  coherent oscillatory behavior in 
the dynamics of the density matrix, which will be further discussed in Sec.~\ref{trans2}. 
%From the real part diagrams presented in the upper part of
%Fig.~\ref{fig2}, we see that except for the case of very small coupling
%constants ($\beta V\ll 1$), the first excited state eigenvalue is not quite
%well separated from the upper eigenvalues so that there are multiple time
%scales involved in the electron transfer kinetics, especially for the symmetric
%case in Fig.~\ref{fig2}(a). In the highly inverted regime case,
%$\beta|\epsilon|=20$ in Fig.~\ref{fig2}(c), the first excited state is quite
%well separated from all the upper states.
In the imaginary part diagram, we find an interesting behavior that imaginary
parts of the eigenvalues asymptotically follow the Rabi oscillation frequency,
\be Z''\approx\omega_{{\rm Rabi}}=\sqrt{4V^2+\epsilon^2}, \label{rabi} \ee
in the strong coupling limit, and this demonstrates {\em the signature of
adiabatic Rabi oscillation picture of the electron transfer in the strong
electronic coupling case}. 

\subsection{Crossover from Nonadiabatic to Adiabatic Regimes}
\label{trans1}

In Fig.~\ref{fig3} we compared the first nonzero eigenvalue of the
nonadiabatic diffusion operator with several predictions from existing theories
in order to characterize various kinetic regimes in the nonadiabatic diffusion
equation.
% as the electronic coupling constant is increased; from the Marcus to
%the solvent-controlled and to the adiabatic regime.
The filled circles are calculation results of the first nonzero eigenvalue
obtained from the spectral analysis method. We compare our eigenvalue solution 
with previous theoretical results, each of which is applicable for different limits;

(i) {\em weak coupling limit} : The eigenvalue solution is compared with
the purely nonadiabatic transition rate without the solvent diffusion effects,
Eq.~(\ref{kNAtot}) (solid line) and with the nonadiabatic-diffusion rate with
the solvent diffusion effects included, Eq.~(\ref{ktot}) (long dashed line).
In any case, we compare the eigenvalue solution with
the sum of the forward and backward rates.
since the eigenvalue solution gives the overall relaxation rate of the density
matrix. 

(ii) {\em strong coupling limit} : We compare the eigenvalue solution with the
adiabatic barrier crossing rates as discussed in Sec.~\ref{adlimit}. We
calculated two different adiabatic reaction rates, the lowest nonzero
eigenvalue given in Eq.~(\ref{eigad}) (dot-dashed line) and the
flux-over-population rate given in Eq.~(\ref{flux}) (short dashed line),
respectively.

%(i) weak couplin limit : 
In the slow bath relaxation case ($\beta\Omega=0.05$) given in
Fig.~\ref{fig3}(a), when the the coupling constant is very small 
($\beta V\lesssim 0.1$)
({\em Marcus regime}), the eigenvalue solution agrees well both 
with the nonadiabatic transition rate 
and with the nonadiabatic-diffusion rate.
The nonadiabatic transition is the rate limiting step in this case.  
As the coupling constant becomes large and so does the nonadiabatic transition rate, 
the overall rate is now affected by the solvent relaxation rate, 
and the eigenvalue solution follows the 
nonadiabatic-diffusion rate, Eq.~(\ref{ktot}), and 
this is the {\em solvent-controlled regime}. 

%(ii) strong coupling limit : 
As the coupling constant becomes 
much larger than the thermal energy, the eigenvalue solution shows a saturation
behavior at $\beta V\sim 1$ first, and starts to increase rapidly as the
electronic coupling becomes stronger, and this indicates the 
{\em adiabatic barrier crossing} in the strong coupling regime. 
In this case, $\beta V\gg 1$, the 
eigenvalue shows a qualitative agreement with 
the {\em adiabatic reaction rates}, Eq.~(\ref{eigad}) and
Eq.~(\ref{flux}).

As we increase the solvent relaxation rate $(\beta\Omega=1)$ in Fig.~\ref{fig3}(b),
%the first thing to notice is that the delocalized transition rate deviates from
%the Marcus result. This was mentioned as the dynamic bath effect in the
%previous section and will be discussed in detail in Sec.~\ref{bath}.
the distinction between the nonadiabatic and the adiabatic regime in the
eigenvalue solution becomes less obvious
than in the slow bath case, Fig.~\ref{fig3}(a).
In the fast bath relaxation case, as the nonadiabatic transition rate becomes large
to be in the solvent-controlled regime, the adiabatic barrier
crossing is already appreciable, so the transition between 
the solvent-controlled 
and adiabatic barrier crossing regimes is not so clear.
This is related to a dynamical modulation of the nonadiabatic transition rate 
in the fast bath case, and will be discussed in details in Sec.~\ref{bath}.

The agreement between the eigenvalue solution and the two adiabatic rate
calculations is qualitative in the strong coupling case. Although it is
reasonable to treat the electron transfer process as an adiabatic barrier crossing reaction on the
lower adiabatic surface when the coupling constant is large 
$(\beta V\gg1)$,
%which rendered Zusman to calculate the adiabatic rate after the diagonalization
%of the electronic part of the Hamiltonian in his original
%work,\cite{zusman-cp-80}
it should be mentioned that a rigorous mathematical proof has not been given to
show that the electron transfer rate from the nonadiabatic diffusion equation
really corresponds to the adiabatic reaction rate in the strong coupling limit.
%for the adiabatic lower surface obtained via the diagonalization of the two
%state Hamiltonian.
Therefore, it may not be surprising that the agreement between the eigenvalue
solution and the adiabatic solutions is only qualitative. Related to this problem,
it should be mentioned that the original nonadiabatic diffusion equation is
based on diabatic representations of electronic states, therefore it may
not give quantitatively the same rate constant as that from the adiabatic
diffusion equation in the strong coupling case.\cite{golosov-jcp-01-1,golosov-jcp-01-2}
%Based on the same physical argument,
%Zusman has calculated the adiabatic reaction rate
%after the diagonalization of the electronic part of Hamiltonian.
%We also note that the barrier height considered in this calculation is not high
%enough. For the symmetric case, both the forward and backward barrier heights
%of the electron transfer reaction are given by $E_b=\lambda/4$, which is
%$2.5k_BT$ in this case. Therefore, before the adiabatic crossing regime appears
%significantly at large coupling constant cases, additional effects such as the
%localization-delocalization transition resulting from the crossing of the
%eigenvalues may affect the first nonzero eigenvalue significantly. We will
%further discuss the localization-delocalization transition phenomenon in
%Sec.~\ref{trans2}. Also, due to not a sufficiently high barrier considered,
%approximate methods for the calculation of the adiabatic rate constant such as
%flux-over-population method, may not be entirely valid. At large reorganization
%energy we have found nonphysical behavior of eigenvalues, and it will be
%discussed later in Sec.~\ref{unphysical}.

The transition from the Marcus to solvent-controlled regimes has been
studied theoretically by using different theoretical methods such as the
projection operator techniques\cite{sparpaglione-jcp-88,yang-jcp-89,cho-jcp-97}
and the path-integral methods.\cite{makarov-cpl-95}
%There are also many experimental
%confirmations in various systems.\cite{weaver-chemrev-92,khoshtariya-jpca-01}
These theoretical approaches do not cover the adiabatic barrier crossing 
regime. 
The crossover behavior from the nonadiabatic to adiabatic
regimes has been observed in quantum mechanical approaches to
electron transfer theories based on the instanton
solution\cite{cao-jcp-95,schwieters-jcp-98,jang-jcp-01} and the diagrammatic
technique\cite{stockburger-jcp-96} which do not take into account the solvent
dynamics effects. The eigenvalue solution presented here clearly demonstrates
transitions between three different regimes.

\subsection{Transition from Incoherent to Coherent Regimes}
\label{trans2} For the symmetric electron transfer reaction case$(\epsilon=0)$
the lower adiabatic surface, $U_{-}$, has a double well structure if
$V<\lambda/2$. When the electronic coupling constant becomes even larger than
$\lambda/2$, the lower adiabatic surface becomes a single well without any
barrier. We consider this situation as a thermodynamic transition from the
localized to delocalized electronic 
states when viewed from a perspective of the
diabatic states. Even though the free energy surfaces of the donor and acceptor
states support the large energy barrier between them in the diabatic picture, a
large coupling constant in the same order of magnitude 
as the reorganization energy makes
the distinction between the donor and acceptor states inappropriate. Instead,
two new single well potential surfaces are obtained, whose eigenfunctions are
neither donor nor acceptor wavefunctions, and 
rather linear combinations of them.
This situation occurs in {\em mixed valence compounds and other strongly coupled systems 
where $ V\sim
\lambda\gg k_B T$}.
\cite{lucke-jcp-97,evans-jcp-98,jung-jpca-99,golosov-jcp-01-1,golosov-jcp-01-2}
The critical value of $V=V_c$ at which the 
double well structure disappears in the lower adiabatic surface, $U_{-}$ 
can be
obtained from Eq.~(\ref{upm}) 
%by identifying the condition of the
%disappearance of the double well structure of $U_{-}$ 
when the reorganization
energy is $\lambda$ and the energy bias is $|\epsilon|$,
\be
(2V_c)^{2/3}+|\epsilon|^{2/3}=\lambda^{2/3}, \label{trans}
\ee
which indicates that the localization-delocalization transition 
occurs at a smaller value of $V_c<\lambda/2$ in the asymmetric reaction case.
Recently, Cukier and
co-workers, and H\"anggi and co-workers have studied such strong coupling
regimes based on the semi-classical approaches.
\cite{casado-jcp-00,casado-cp-01,hartmann-jcp-00,goychuk-cp-01}

%\begin{figure}
%\bc
%\epsfxsize=3 in
%\rotate{\rotate{\rotate{\epsffile{\figdir/fig4.eps}}}}
%\caption{
%Transition between the incoherent and the coherent regimes in the symmetric electron transfer case
%is manifested in the calculation of the first and higher excited state eigenvalues.
%When the coupling constant is about half of the reorganization energy,($\beta\lambda=10$), the
%higher excited states cross with the first excited state. The symbols which are on top of each other
%are the same real parts of the complex conjugate eigenvalues.
%Other parameters are $\beta\Omega=1$ and $\beta\epsilon=0$.
%}\label{fig4}
%\ec
%\end{figure}

To investigate the feature of the localization-delocalization transition in the
spectral structure, we show the lowest 5 
eigenvalues as a function of the coupling
constant in the strong coupling regime when $\epsilon=0$ 
in Fig.~\ref{fig4}. When the 
electronic coupling is about the half of the reorganization energy as 
predicted by Eq.~(\ref{trans}) 
$\beta V\gtrsim 5$, 
the first nonzero
eigenvalue (solid line) 
is no longer well separated from higher complex eigenvalues, 
it is no longer valid to 
describe the dynamics of the density matrix by a incoherent 
rate process. Also, there is a rapid drop of the first excited eigenvalue
around $\beta V\sim 6$ due to the crossing of the first excited state
eigenvalue and higher complex eigenvalues 
(note that $\beta V_c=5$ in this case). 
{\em The 
crossing behavior of the first and higher excited state eigenvalues is a
signature of the transition between the incoherent to coherent
regimes in electron transfer processes}.

A crossing of the two lowest nonzero eigenvalues can be used as 
an approximate criterion for the transition from the incoherent to coherent regimes. 
In Fig.~(\ref{phase}), we plotted the value of the electronic coupling constant, $V_c$,  
where a crossing of the two lowest nonzero eigenvalues 
occur for a given energy bias, $|\epsilon|$, and the prediction of Eq.~(\ref{trans})   
for the case $\beta\Omega=0.01$ and $\beta\lambda=10$. 
They show the same trend that the coherent regime appears earlier in 
the asymmetric reaction case than in the symmetric reaction case. 

%Moreover, the second and the third lowest
%eigenvalues become complex conjugate to each other, lying quite close to the
%first excited state eigenvalue.
%In other words, the electron transfer ``rate'' is not sensibly defined any more
%in such a large coupling constant case, and the temporal evolution of the
%density matrix will show the oscillation behavior as well as the
%multi-exponential decay.
%The electron transfer ``rate'' is not sensibly defined any more in such a large
%coupling constant case, and the temporal evolution of the density matrix will
%show the oscillation behavior as well as the multi-exponential decay

\subsection{Dynamic Bath Effect}
\label{bath}
%As observed in Fig.~\ref{fig3}(b) the delocalized nonadiabatic transition rate shows a deviation
%from its static limit (Marcus result) when the solvent relaxation rate is fast, typically  $\beta\Omega\approx 1$.
%Effects of the bath relaxation dynamics appear as two-fold in the overall
%electron transfer rate in the nonadiabatic regime. First, as discussed before,
%the overall electron transfer rate is a combined result of 
%the nonadiabatic transition and solvent diffusion processes, and the finite solvent relaxation rate affects
%the overall rate in Eq.~(\ref{ktot}) in the solvent-controlled limit.
%The adiabaticity parameter is defined as the ratio
%of the nonadiabatic transition rate to the solvent relaxation rate to quantify the
%kinetic bath effect,\cite{zusman-cp-80,garg-jcp-85}
%%\be \xi={2\pi V^2\over\Omega\lambda}
%\label{xi}. \ee
%%
In addition to the fact that the overall electron transfer rate appears a combined result of 
the nonadiabatic transition and solvent diffusion processes 
% and the finite solvent relaxation rate affects
%the overall rate 
in Eq.~(\ref{ktot}),
% in the solvent-controlled limit.
%that yielding the overall reaction rate 
the nonadiabatic transition rate itself, Eq.~(\ref{kNA}), is
modulated by the dynamic nature of the solvent relaxation process 
that is 
characterized 
by the bath correlation
function in Eq.~(\ref{g}). We will call this a {\it
dynamic bath effect}. This will be evident if the bath relaxation rate  
is comparable to the amount of the polarization energy fluctuation, 
\be \Omega \sim \Delta=\sqrt{2\lambda k_B T}. 
\ee

To investigate the dynamic bath effect, we compare the eigenvalue solution and
two different reaction rates, Marcus rate in Eq.~(\ref{kmartot}) (static bath limit)
and the nonadiabatic transition rate, Eq.~(\ref{kNAtot}) (dynamic
bath effect included) in the weak coupling limit as we vary the energy
bias in Fig.~\ref{fig5}.
%
% in the slow bath dynamics ($\beta \Omega=0.1$) and in the fast bath
%dynamics case ($\beta \Omega=1$).
%
% (a) slow bath case($\beta\Omega=1$,$\beta V=1/10$), (b)
%slow bath, weak coupling case($\beta\Omega=1/9$,$\beta V=1/30$), (c) fast bath,
%strong coupling case($\beta\Omega=1$,$\beta V=1/2$), (d) slow bath, strong
%coupling case($\beta\Omega=1/9$,$\beta V=1/6$). To estimate the dynamic bath
%effect separately from the kinetic bath effect, we have kept the adiabaticity
%parameter constant when comparisons between the slow and fast bath cases in the
%weak ($\xi={2\pi \over 100}$) and strong ($\xi={2\pi \over 4}$) coupling cases.
%
%\newpage
%\begin{figure}
%\epsfxsize=2.7 in
%\bc
%(a) \\
%\vspace{-1cm}
%\rotate{\rotate{\rotate{\epsffile{\figdir/fig5a.eps}}}}
%(b) \\
%\vspace{-1cm}
%\rotate{\rotate{\rotate{\epsffile{\figdir/fig5b.eps}}}}
%(c) \\
%\vspace{-1cm}
%\rotate{\rotate{\rotate{\epsffile{\figdir/fig5c.eps}}}}
%\rule{0cm}{4cm} \\
%(d) \\
%\vspace{-1cm}
%\rotate{\rotate{\rotate{\epsffile{\figdir/fig5d.eps}}}}
%\caption{
%The eigenvalue solution shows the dynamical bath effect
%for the fast bath relaxation case shown in
%Figs. 5(c) and (d)
%($\beta\Omega=1$) while it converges to the static result
%for the slow relaxation case ($\beta\Omega=1/9$) shown in 5(a) and (b).
%The adiabaticity parameters are set to equal in the fast and slow bath relaxation cases
%to ensure that the peak shift is due to the dynamical nature of bath.
%The reorganization energy is $\beta\lambda=10$ in all cases.
%}
%\labe{fig5}
%\ec
%\end{figure}
%
All the calculation results show a characteristic behavior of the Marcus
curve; they show a maximum value when $|\epsilon|\sim \lambda$.
%The reason
%that the curves appear asymmetric with respect to the maximum position is due
%to the Boltzmann factor involved when we plot the overall relaxation rate.
In the slow bath case, Fig.~\ref{fig5}(a), all the three curves agree with each
other since the dynamic bath effect is negligible in this case. In the fast
bath case, Fig.~\ref{fig5}(b), we notice that 
the eigenvalue and the nonadiabatic transition rate 
agree with each other, and show a deviation from the Marcus rate.  
In particular, in the fast bath case, {\em the maximum position of 
the eigenvalue as well as of the nonadiabatic transition rate shift 
toward a smaller value of the energy bias compared with the Marcus rate}.
%This is {\em the dynamic bath effect}.

In the static bath limit, 
the nonadiabatic transition rate is reduced to the Marcus rate, exhibiting 
the maximum position at the activationless case, $|\epsilon|=\lambda$. 
%When the solvent relaxation rate is slow, the solvent reorganization energy is
%given by its static value; however, 
When the bath relaxation is fast, the
solvent reorganization energy is effectively reduced, and the maximum electron
transfer rate appears at a smaller value of the bare solvent reorganization
energy, exhibiting the peak position in the normal regime, $|\epsilon| < \lambda$,
and this results in a significant deviation of the nonadiabatic transition 
rate from the Marcus rate. 
%In the strong coupling case,
%Figs.~\ref{fig5}(c) and (d), the dynamic bath effect is still present as can be
%seen from the peak shift from its static limit. In this case, the eigenvalue
%solution is greater than the nonadiabatic-diffusion rate due to the adiabatic
%
%In order to estimate the amount of the peak shift as a function of the bath
%relaxation rate, we calculated the nonadiabatic transition rate as a function
%of the energy bias for several values of the solvent relaxation rates and compared
%them with its static limit given by the Marcus formula, Eq.~(\ref{k12mar}) in
%Fig.~(\ref{fig6})(a).
%When the bath is sluggish enough ($\beta\Omega\le0.1$) the
%maximum of the nonadiabatic transition rate occurs at $-\epsilon=\lambda$ as
%the Marcus formula predicts; however, as the bath relaxation becomes faster,
%the value of the energy gap which shows the maximum transition rate shifts into
%a smaller value.
The amount of the peak shift as a function of the solvent relaxation rate can also be 
estimated from Eq.~(\ref{kNA}), and it is shown in Fig.~\ref{fig6} that 
the peak shift increases linearly with $\Omega$.
%when $\beta \Omega \le 1$ and saturates to the reorganization energy value at
%large $\Omega$.
%\rule{0cm}{3cm}
%\begin{figure}
%\bc
%\epsfxsize=2.7 in
%{\epsffile{\figdir/fig6.eps}}
%\end{center}
%\caption{
%The dynamical bath effect is estimated through the calculation of the forward
%delocalized nonadiabatic transition rate as a function of the energy bias for various
%solvent relaxation rates in Fig.~\ref{fig6}(a). When the bath is static, the nonadiabatic transition rate
%shows a maximum at $\epsilon=\lambda$. As the bath relaxation increases, the maximum shifts
%into the smaller value of the energy bias. In Fig.~\ref{fig6}(b)
%the shift of maximum of the forward delocalized transition rate from its
%static limit is plotted as a function of the solvent relaxation rate.
%}\label{fig6}
%\end{figure}

The peak shift observed in the eigenvalue solution has a close analogy to the
Stokes shift observed in the condensed phase
spectroscopy.\cite{mukamel-spec-95} It is worthwhile to mention that $g(t)$
obtained in Eq.~(\ref{g}) has exactly the same form as the line broadening
function obtained in the hight temperature limit of the overdamped Brownian oscillator model 
in the condensed phase spectroscopy,\cite{yan-jpc-88,mukamel-spec-95} and the nonadiabatic transition rates
given in Eq.~(\ref{kNA}) have the same functional forms as the line shapes of
the chromophore interacting with the Debye solvent. As discussed in the
App.~\ref{app-nonad}, the Stokes shift between the absorption and the emission profiles
disappears in the fast bath case 
%(also called the motional narrowing phenomenon 
%in the lineshape theory) 
and it has the same origin as the peak shift
observed here.

Implication of the dynamic bath effect on the experimental observables may be
significant. The reorganization energy of the solvent can be estimated from the
maximum of the electron transfer rates based on a series of the measurements
of electron transfer rates for different systems of which the energy bias
and the electronic coupling constants are known. 
However, the dynamic bath effect can make the value of the
reorganization energy measured in the experiment smaller than its true value.
Therefore, care should be taken to disentangle the dynamical bath effect from
the measured value of the solvent reorganization energy. 
%This may be done by a separate
%experiment which gives the solvent relaxation rate independently.

\subsection{Density Matrix Propagation}
\label{dynamics} The spectral method can be used as a propagation method of the
density matrix as shown in Eq.~(\ref{22}). We demonstrate the time evolution of
the donor and acceptor populations, $P_1(t)$ and $P_2(t)$ based on the spectral
method. Parameters are chosen as $\beta\lambda=10$, $\beta\Omega=1$, and
$\beta\epsilon=0$. The electronic coupling constant will be varied.
%
%The donor and acceptor populations, $P_1(t)$ and $P_2(t)$, are defined as the
%spatial averages of the diagonal elements of the density operator at time $t$,
%\be
%P_D(t)&=&\int_{-\infty}^{\infty}dE\rho_{11}(E,t), \no  \\
%_A(t)&=&\int_{-\infty}^{\infty}dE\rho_{22}(E,t).
%\ee
%
The initial condition is chosen such that at $t=0$ only the donor state is
populated with the equilibrium distribution in the diabatic state, \be
\rho_{11}(E,0)&=&\rho_{11}^{\rm eq}(E),  \\
\rho_{22}(E,0)&=&\rho_{12}(E,0)=\rho_{21}(E,0)=0. \ee
This initial condition corresponds to that of the photo-induced back electron transfer
reaction. \cite{evans-jcp-98,jung-jpca-99,golosov-jcp-01-2,reid-jpc-95}
%\begin{figure}
%\bc
%\epsfxsize=2.7 in
%\rotate{\rotate{\rotate{\epsffile{\figdir/fig7.eps}}}}
%\ec
%\caption{
%Invalidity of the nonadiabatic diffusion equation
%at large reorganization energy ($\beta\lambda\gg 1$).
%Real parts of the lowest eigenvalues(${\rm Re}Z$) show a negative value
%in the case of large reorganization energy ($\beta\lambda\ge15$).
%Other relevant parameters are $\beta V=0.1$ and $\beta\Omega=1$.
%}
%\label{fig7}
%\end{figure}

In Fig.~\ref{fig7} we show the time evolution of the donor (solid line) and the
acceptor (dashed line) state populations with various electronic coupling
constants. At the smallest coupling constant, $\beta V=0.5$, the donor and
the acceptor populations relax exponentially to the equilibrium at long times. The
electronic coupling constant is small such that the time evolution is governed
by the lowest nonzero eigenvalue of the nonadiabatic diffusion equation. As the
electronic coupling constant is increased, the temporal behavior of 
the density matrix undergoes the
incoherent-coherent transition as discussed based on the spectral analysis in
Sec.~\ref{bath}. When the coherent oscillation is observed in the temporal
evolution of the density matrix($\beta V=2$ and $\beta V=4$), the oscillation
frequency matches the Rabi oscillation frequency given in Eq.~(\ref{rabi}).
%
%This demonstrates the ability to propagate the density matrix based on the
%spectral analysis and confirms some predictions on the temporal evolution of
%the density matrix studied in previous sections based on the spectral analysis
%method.
%density such as
%the exponential decay at weak coupling, the adiabatic Rabi oscillation at
%strong coupling, and moreover, the incoherent-coherent transition.
The time-dependent study of the nonadiabatic diffusion equation based on the
spectral method was also performed for back electron transfer dynamics occurring in
the mixed-valence system.\cite{jung-jpca-99}
% when 
%both the electronic coupling 
%%constant and the energy bias are the same order of magnitude 
%In that system, the oscillation behavior in the temporal evolution of
%he density matrix has been confirmed via the density matrix propagation
%method, and the Rabi oscillation frequency and the decay rate correlate well
%with the result of the spectra.
%This is a clear indication of the adiabatic picture in the electron transfer
%processes in the strong coupling case.

\subsection{Nonphysical Behavior in Nonadiabatic Diffusion Equation}
\label{unphysical} We found that the nonadiabatic diffusion equation can yield
nonphysical behavior in some parameter regimes. We checked the case of large
reorganization energy, $\beta\lambda \gg 1$, and found that in this case the
smallest eigenvalue can be negative instead of zero,
which yields an exponential growth of the density matrix
in the long time limit. Fig.~\ref{fig8} demonstrates how two lowest real
eigenvalues change as the reorganization energy is varied in the weak coupling
case. When $\beta\lambda \gtrsim 12$, the lowest nonzero eigenvalue becomes
negative instead of positive, thus violating the positive definiteness of the
density matrix  in the nonadiabatic diffusion equation.

%\begin{figure}
%\begin{center}
%\epsfxsize=2.5 in
%\epsffile{\figdir/fig8.eps}
%\caption{
%Invalidity of the nonadiabatic diffusion equation
%at large reorganization energy ($\beta\lambda\gg 1$).
%Real parts of the lowest eigenvalues(${\rm Re}Z$) show a negative value
%in the case of large reorganization energy ($\beta\lambda\ge15$).
%Other relevant parameters are $\beta V=0.1$ and $\beta\Omega=1$.
%}
%\label{fig8}
%\end{center}
%\end{figure}

%This manifests problematic behaviors of the nonadiabatic diffusion equation in
%the regime where $\lambda\gg k_BT$. 
There are several possible reasons for this problematic behavior. 
One of them is an intrinsic limitation in consistent treatments of
the electronic and nuclear degrees of freedom in the mixed quantum-classical
approach.
%adapted in the nonadiabatic diffusion equation.
%and the fact that the nuclear dynamics is
%treated as purely classically and the electronic transition as purely quantum
%mechanical transition.
Similar problem in the mixed quantum-classical approach has been pointed out in
other contexts.\cite{bader-jcp-94,egorov-jpcb-99} The other possible reason for
the positivity violation may be the Markovian approximation used in the
calculation of the frictional kernel in the overdamped regime for {\em the
nonadiabatic dynamics}. Although the Markovian approximation for the friction
kernel has been successfully applied to the {\em adiabatic} reaction rate
theory where the single potential energy surface is
involved,\cite{hanggi-rmp-90} there have been limited studies on the validity
of this approximation in the {\rm nonadiabatic} reaction
cases.\cite{roy-jcp-94,tang-jcp-96} Along this line, similar nonphysical
behaviors in the nonadiabatic diffusion equation yielding negative electron
transfer rates have been found
recently.\cite{frantsuzov-cpl-97,frantsuzov-jcp-99,thoss-jcp-01}
% and it has
%been attributed to the Markovian approximation involved in the derivation of
%the nonadiabatic diffusion equation in .
%from the microscopic quantum mechanical model, which may
%not be well justified in the electron transfer problem.
The precise origin of this artifact is still not clear, and more studies need
to be done to clarify this issue in a decisive way.

\section{Conclusion}
\label{concl} In this paper we studied the electron transfer reaction
kinetics in Debye solvents described by the two-state nonadiabatic diffusion
equation. By applying the spectral analysis method developed recently, we 
investigated various kinetic regimes of the electron transfer in Debye solvents 
in a unified way. 
The results obtained in this work are summarized;

(i) The general spectral structure of the nonadiabatic diffusion equation has
been revealed. It exhibits a tree-like structure with three major branches; 
a single branch of real eigenvalues corresponding to multi-exponential decays and
two symmetric branches of complex conjugate eigenvalues corresponding to damped
oscillations.

(ii) Several kinetic regimes in the electron transfer reactions in solutions 
have been identified. 
The first nonzero eigenvalue solution bridges between the Marcus, the
solvent-controlled, 
and the adiabatic crossing regime in the incoherent kinetics 
regime. It agrees well with nonadiabatic reaction rates 
from previous theoretical results 
in the weak coupling regime. 
In the adiabatic crossing regime the eigenvalue solution agrees with 
adiabatic reaction rates qualitatively.

(iii) When the electronic coupling constant is about half the reorganization
energy in the symmetric reaction case (less than half in the asymmetric reaction case), the
eigenvalue diagram shows coalescence/bifurcation behavior, where the two lowest
nonzero eigenvalues cross each other and become complex conjugate. This
indicates the damped oscillation behavior in the overdamped solvent, and
correlates with the localization-delocalization transition in the lower
adiabatic surface.

(iv) Due to the finite timescale of the solvation, the eigenvalue solution 
is modulated in the weak coupling regime such that the maximum position 
of the electron transfer rate appears at a smaller value of the bare solvent 
reorganization energy in the Marcus curve, which shows a substantial deviation 
from the standard Marcus theory. 
The consequence of the dynamic bath effect  may be significant
in the experimental determination of the solvent reorganization energy.

(v) The spectral analysis method was used as a propagation scheme to study the
time-dependent behavior of the density matrix, and it has confirmed previous
predictions made based on the investigation of the spectral structure.

(vi) Nonphysical behavior of the nonadiabatic diffusion equation, violating the
positivity, has been identified in the large reorganization energy case by the
spectral analysis method, and the precise origin of this problem should deserve
further investigations.

In the isolated quantum system the eigenvalue solution method offers a powerful
way to analyze the dynamics. In comparison, the spectral structure of the
dissipative quantum dynamics has not been fully explored yet, and most studies
on the quantum dissipative system employ trajectory methods. In this regard,
our studies on the spectral structure of the electron transfer kinetics can be
a good motivation to alternative studies of various dissipative systems. The
nonadiabatic diffusion equation studied in this work is based on the mixed
quantum-classical approach. However, the spectral analysis method itself can be
applied to other quantum dissipative dynamical equations in general such as the
Redfield equation,\cite{jean-jcp-92,cao-jcp-97a} quantum Fokker-Planck
equation,\cite{tanimura-pra-91,yan-pra-98} and quantum master
equation,\cite{caldeira-physa-83} and the studies based upon the spectral
analysis will bring more direct insights on the dissipative dynamics.

\section*{Acknowledgments}
YJ thanks Dr.~Seogjoo Jang and  Prof.~Robert J.~Silbey for useful discussions
and Prof.~Robert J.~Silbey for his support.
This work was supported by
grants from Research Corporation and National Science Foundation.

\appendix
\section{Spectral analysis for Fokker-Planck operator on a single 
harmonic potential}
\label{app-single}
For a single harmonic potential, $U(x)=\frac{1}{2} m\omega^2 x^2$, 
the Fokker-Planck operator is given by  
\be
\cl_{\rm FP}=D\left({{\p^2}\over{\p x^2}}+{\beta {\p \over {\p x}} U'}\right), 
\ee
and it is possible to transfrom $\cl_{\rm FP}$ into a quantum mechanical Hamiltonian corresponding to 
a fictitious harmonic oscillator 
in imaginary time,
\be H =-e^{\beta U(x)/2}\cl_{\rm FP}e^{-\beta U(x)/ 2}
    =-{1 \over {2\mu}} {\p ^2 \over {\p x^2}} + V(x), \label{11}
\ee
where $\mu=(2D)^{-1}$ and the potential for the fictitious harmonic oscillator is
given by 
\be V(x) = D \left[{1 \over 4}
(\beta U^{'}(x))^2 -{1 \over 2} \beta U^{''}(x) \right]
    = {1 \over 2} {\mu \gamma^2 x^2} - {\gamma \over 2}, \label{12}
\ee
with $\gamma=D m \omega^2/k_BT$. Since the transformed potential in
Eq.~(\ref{12}) is in the same form as that of a harmonic oscillator
model with an off-set of the zero point energy, eigenvalues and 
eigenfunctions for the original Fokker-Planck operator can be obtained 
immediately from the eigenvalue solutions of the  harmonic oscillator
Hamiltonian,\cite{risken-fpe-84}
\be
\cl_{\rm FP}|\psi_n^{R}\ra=-n\gamma|\psi_n^{R}\ra, \\ 
\la \psi_n^{L}|\cl_{\rm FP}=-n\gamma\la \psi_n^{L}|,
\ee
where $\psi_{n}^{L}(x)=H_n(x/\sqrt{2}x_0)
/((2\pi)^{1/2} 2^n n! x_0)^{1/2}$ and $\psi_{n}^{R}(x)=e^{-x^2/2x_0^2}\psi_{n}^{L}(x)$ with 
$x_0=\sqrt{k_BT/m\omega^2}$ and $H_{n}$ being the $n$th order Hermite polynomial.

\section{Rate Equations in Nonadiabatic Regime}
\label{app-nonad}
We briefly review 
the rate expressions in the nonadiabatic limit 
($\beta V \ll 1$). 
\cite{sparpaglione-jcp-88,yang-jcp-89}
First, we obtain a formal solution for $\rho_{12}(E,t)$ in the time domain from Eq.~(\ref{2c}),
\be
\rho_{12}(E,t)=iV\int_0^t {\rm d}\tau\int_{-\infty}^{\infty} {\rm
d}E_0G_{12}(E,\tau|E_0)
                     \left[\rho_{11}(E_0,t-\tau)-\rho_{22}(E_0,t-\tau)\right],  \label{rho12_1}
\ee
where $G_{12}(E,t|E_0)$ is the Green's function for the operator, 
${\cl_{12}}-i\omega_{12}$,
\be {G}_{12}(E,t|E_0)=\la E | e^{(\cl_{12}-i\omega_{12})t}| E_0\ra.
\label{g12}
\ee
%
%
%At first it is convenient to take the Laplace transformation of the equation
%for $\rho_{12}(E,t)$ in Eq.~(\ref{2c}), \be
%\hat{\rho}_{12}(E,s)=iV\int_{-\infty}^{\infty}{\rm d}E_0\hat{G}_{12}(E,s|E_0)
%                              [\hat{\rho}_{11}(E_0,s)-\hat{\rho}_{22}(E_0,s)], \label{rho12}
%\ee where the initial condition $\rho_{12}(E,0)=0$ is used, and
%$\hat{G}_{12}(E,s|E_0)$ is the Green's function for $\hat{\rho}_{12}(E,s|E_0)$
%satisfying
%%
%\be
%\hat{G}_{12}(E,s|E_0)=(s-\cl_{12}+i\omega_{12})^{-1}\delta(E-E_0). \label{g12}
%\ee
%%
%Then by taking the inverse Laplace transform of Eq.~(\ref{rho12}), we
%btain the formal solution for $\rho_{12}(E,t)$ in the time domain
%\be
%\rho_{12}(E,t)=iV\int_0^t {\rm d}\tau\int_{-\infty}^{\infty} {\rm d}E_0G_{12}(E,\tau|E_0)
%                     \left[\rho_{11}(E_0,t-\tau)-\rho_{22}(E_0,t-\tau)\right] \label{rho12_1}
%\ee
Substituting $\rho_{12}$ and $\rho_{21}=\rho^*_{12}$ from
Eq.~(\ref{rho12_1}) into the equations for $\rho_{11}$ and $\rho_{22}$ in
Eqs.~(\ref{2a}) and~(\ref{2b}), we can obtain a closed set of
integro-differential equations for $\rho_{11}$ and $\rho_{22}$,
\be
{\p\over\p t}\rho_{ii}(E,t)&=&\cl_{ii}\rho_{ii}(E,t) \no \\
                                      &-&2V^2{\rm Re}\int_{-\infty}^{\infty}{\rm d}E_0
                                       \int_{0}^{t}{\rm d}\tau G_{12}(E,\tau|E_0)
                                       [\rho_{ii}(E_0,t-\tau)-\rho_{jj}(E_0,t-\tau)]. \label{dr1dt}
%{\p\over\p t}\rho_{22}(E,t)&=&\cl_{22}\rho_{22}(E,t) \no \\
%                                      &+&2V^2{\rm Re}\int_{-\infty}^{\infty}dE_0
%                                        \int_{0}^{t}d\tau G_{12}(E,\tau|E_0) \no \\
%                                      &\times&[\rho_{11}(E_0,t-\tau)-\rho_{22}(E_0,t-\tau)].  \label{dr2dt}
\ee

%So far we have not made any approximation in solving the original nonadiabatic
%diffusion equation but have formally reduced four component equations into two
%component equations which involve the convolution integrals.
%When the electronic coupling constant is very small compared with the thermal
%energy
In the weak coupling regime, $\beta\D\ll 1$, the electron transfer process is
characterized as nonadiabatic, and the transition between 
two electronic states takes place
only when the bath polarization energy is close to the free energy gap, 
and $\rho_{11}$ and $\rho_{22}$ vary slowly compared with $\rho_{12}$ and $\rho_{21}$. 
We can assume that the coherent Green's function $G_{12}$
varies much faster than the population difference $\rho_{11}-\rho_{22}$ in this
case and that dominant contribution to the integral in Eq.~(\ref{dr1dt}) comes from
the short time region. 
Using these arguments we can approximately deconvolute the integrals and
extend the upper limit of the integral to the infinity,
\be
&&2V^2{\rm Re}\int_{-\infty}^{\infty}{\rm d}E_0\int_{0}^{t}{\rm d}\tau G_{12}(E,\tau|E_0)
[\rho_{11}(E_0,t-\tau)-\rho_{22}(E_0,t-\tau)] \no \\
&\approx&K(E)[\rho_{11}(E,t)-\rho_{22}(E,t)],
\ee
where $K(E)$ is the rate kernel 
and is given in terms of the coherent Green function
$G(E,t|E_0)$ defined in Eq.~(\ref{g12}),\cite{sparpaglione-jcp-88,yang-jcp-89}
\be K(E)=2V^2{\rm Re}\int_0^{\infty}{\rm d}\tau\int_{-\infty}^{\infty}{\rm
d}E_0 G_{12}(E,\tau|E_0). \label{ke} \ee
%
%which leads to Eq.~(\ref{red1}) with the coordinate dependent transition rate 
%defined Eq.~(\ref{ke}). 
%%
%\be K(E)=2V^2{\rm
%Re}\int_0^{\infty}d\tau\int_{-\infty}^{\infty}dE_0 G_{12}(E,\tau|E_0).
%%%
%
The approximations made here are well established in the normal regime of the
electron transfer, but they are questionable in the inverted or activationless
regime, and improvements on these regimes have been made. \cite{rips-jcp-87b}
%\subsection{Nonadiabatic Rates} \label{narate}

The nonadiabatic rates are given in terms of the coherent Green's
function,\cite{sparpaglione-jcp-88,yang-jcp-89}
\be k_{\rm NA}^{i}=2V^2 {\rm Re}\int_0^{\infty}{\rm
d}t\int_{-\infty}^{\infty}{\rm d}E\int_{-\infty}^{\infty}{\rm d}E_0
G_{ij}(E,t|E_0)\rho_{ii}^{\rm eq}(E_0). 
 \label{kNA-app} \ee
The coherent Green's function, $G_{12}(E,t|E_0)$, can be evaluated first by
transforming the Fokker-Planck operator into the harmonic oscillator
Hamiltonian as done in Eq.~(\ref{11}),
\be
H_{12}(E)&=&-e^{\beta {\overline U}(E)/2}(\cl_{12}(E)-i\omega_{12}(E))e^{-\beta {\overline U}(E)/ 2} \no\\
           &=&-{1\over 2m}{\p^2\over\p \tilde{E}^2}+{1\over 2}m\Omega^2 \tilde{E}^2-i\tilde{\epsilon}, \label{H12}
\ee
where $m=1/(2D_E)=1/(2\Omega \Delta^2)$. Two complex energy variables are defined 
by $\tilde{E}=E +i \left({2\Delta^2\over\Omega}\right)$ 
and $\tilde{\epsilon}=\epsilon +i\left({\Delta^2\over\Omega}-{\Omega\over 2}\right)$. 
Since $H_{12}$ is a
simple harmonic oscillator Hamiltonian with respect to the complex energy
variable $\tilde{E}$, we can calculate its
propagator, $\la \tilde{E} | e^{-H_{12}t} | \tilde{E}_{0}\ra$.\cite{feynmann-statmech-72}
%
%\be \la \tilde{E} | e^{-H_{12}t} | \tilde{E}_{0} \ra={1\over\sqrt{4\pi\Delta^2
%\sinh(\Omega t)}}
% \exp\left[-{1\over{4\Delta^2\sinh(\Omega t)}}
%\left((\tilde{E}^2+\tilde{E}_{0}^2)\cosh(\Omega
%t)-2\tilde{E}\tilde{E}_{0}\right)+i\tilde{\epsilon}t\right].
%\no \\
%\label{prop} \ee
%
%Then, $G_{12}$ is given by
%%
%\be
%G_{12}(E,t|E_0)&=&\la E | e^{(\cl_{12}-i\omega_{12})t}| E_0\ra  \no \\
%           &=&e^{-\beta[{\overline U}(E)-{\overline U}(E_0)]/2} \la \tilde{E} | e^{-H_{12}t} | \tilde{E_0}\ra,
%\label{G12} \\
%G_{21}(E,t|E_0)&=&G^*_{12}(E,t|E_0). \ee
%%
%Noting that $\omega_{12}(E)=E-\epsilon$ is a linear function of $E$,
%$L_{12}-i\omega_{12}$ can be cast into the Fokker-Planck operator for the
%harmonic oscillator with displacement in the imaginary axis if we make a
%similar transformation given in Eq.~(\ref{11}),
Then, we can express 
$G_{12}(E,t|E_0)$ in Eq.~(\ref{g12}) 
explicitly,
\be G_{12}(E,t|E_0) 
%&=&e^{-\beta[{\overline U}(E)-{\overline U}(E_0)]/2}
%\la \tilde{E} |e^{-H_{12} t}| \tilde{E_0}\ra \no \\
&=&{1\over\sqrt{4\pi\Delta^2 \sinh(\Omega t)}} \exp\left[-{1\over 4\Delta^2
\sinh(\Omega t)}\left(e^{\Omega t}E^2-e^{-\Omega t}E_0^2-2E E_0\right)
\right.  \no \\
&& \mbox{\makebox[0.5 in]{ }}\left. +\left( {2\Delta^2\over\Omega^2} -
i{(E+E_0) \over\Omega}\right) \tanh\left({\Omega t\over 2}\right)
+i\tilde{\epsilon}t \right]. \label{G12a} \ee
Straightforward Gaussian integrations over the space coordinates in
Eq.~(\ref{kNA-app}) result in 
%given in 
Eq.~(\ref{kNA}).

%The time scale of the nuclear dynamics is typically characterized by the Debye
%frequency, $\Omega$, while the magnitude of the thermal fluctuation is
%proportional to the root-mean-square fluctuation energy,
%$\Delta=\sqrt{2k_BT\lambda}$.
Depending on the timescale of the bath, $\Omega$, and 
the amount of the polarization energy fluctuation, $\Delta=\sqrt{2\lambda k_B T}$, 
the nonadiabatic reaction rates in Eq.~(\ref{kNA}) have two different limits. 
%we have two different kinetic regimes. 
%characterized by a dimensionless parameter 
%$\kappa=\Omega/\Delta$.
\cite{mukamel-spec-95}
In the static bath limit $(\Omega \ll \Delta)$,
%such that the dominant contribution to the integral comes from the region $t\ll\Omega^{-1}$,
we take the short time limit $\Omega t\ll 1$ of $g(t)$ in Eq.~(\ref{g}), 
$g(t)\approx{1\over 2}{\Delta^2 t^2}$,   
then the nonadiabatic transition rate 
yields the Marcus rate given in Eq.~(\ref{kmar}) after a Gaussian integration. 
%%
%\be k_{\rm NA}^{i}&\approx& k_{\rm GR}^{i}
%%=2\pi V^2 \rho_{ii}^{\rm eq}(\epsilon)
%=V^2\sqrt{\pi\over{\lambda k_BT}}
%\exp\left[-{(\epsilon\pm\lambda)^2\over{4\lambda k_BT}}\right], \\
%k_{\rm NA}&\approx& k_{\rm GR}^{1}+k_{\rm GR}^{2} \label{k12mar} \ee
%
%where the sign is chose as $+$ when $ij=12$ and $-$ when $ij=21$.
%k_{21}&\approx& 2\pi V^2 \rho_{22}^{eq}(\epsilon)=V^2\sqrt{\pi\over{\lambda
%k_BT}} \exp\left[-{(\epsilon-\lambda)^2\over{4\lambda k_BT}}\right].
%\label{k21mar} \ee
The maxima of $k_{\rm NA}^{1}$ and $k_{\rm NA}^{2}$ are separated by $2\lambda$
in this case and this is analogous to the Stokes shift observed in condensed
phase spectroscopy.\cite{yan-jpc-88,mukamel-spec-95} When the dynamics of
bath degrees of freedom is very fast $(\Omega \gg \Delta)$, 
we can take the long time limit $\Omega t \gg 1$ of $g(t)$, $g(t)\approx
(\Delta^2/\Omega-i\lambda)t$, which leads to, 
\be k_{\rm NA}^{1}\approx k_{\rm NA}^{2}\approx
{2V^2\gamma\over{\epsilon^2+\gamma^2}}, \label{narrow} \ee
where $\gamma=\Delta^2/\Omega$. The Stokes shift between $k_{\rm NA}^{1}$ and
$k_{\rm NA}^{2}$ has now disappeared as a result of 
fast fluctuation of the solvent polarization energy. 
%The dynamical effect of the bath on the electron transfer rate will
%be discussed later in detail.

%\subsection{Solvent Relaxation Rates} \label{bathrate}
The solvent diffusion rate is given in terms of 
the population Green's function $G_{ii}(E,t|E_0)$ at the crossing point
$E=E_0=\epsilon$ in Eq.~(\ref{kd1}).  
Since $\cl_{ii}$ is given by the Fokker-Planck operator with
a displaced harmonic potential for each electronic state, the population
Green's functions $G_{ii}$ are easily obtained 
%given by the transition probability functions
%for the Ornstein-Uhlenbeck process,
\cite{feynmann-statmech-72,risken-fpe-84}
\be
G_{ii}(E,t|E_0)&=&{1\over\sqrt{2\pi\Delta^2(1-e^{-2\Omega t})}}
\exp\left[-{(E\pm\lambda-(E_0\pm\lambda) e^{-\Omega
t})^2\over{2\Delta^2(1-e^{-2\Omega t})}}\right],
\ee
with $+(-)$ sign for $i=1(2)$, and the solvent diffusion rates can be calculated.  
%
%G_{22}(E,t|E_0)&=&{1\over\sqrt{2\pi\Delta^2(1-e^{-2\Omega t})}} \no \\
%&\times&\exp\left[-{(E-\lambda-(E_0-\lambda) e^{-\Omega
%t})^2\over{2\Delta^2(1-e^{-2\Omega t})}}\right], \ee
%which at $E=E_0=\epsilon$ reduce to
%
%\be
%G_{ii}(\epsilon,t|\epsilon)&=&{1\over\sqrt{2\pi\Delta^2(1-e^{-2\Omega t})}} \no \\
%&\times&\exp\left[-{(E\pm\epsilon)^2(1-e^{-\Omega t})^2\over{2\Delta^2(1-e^{-2\Omega t})}}\right],   \label{g1c}
%\ee
%%
%%\\
%%G_{22}(\epsilon,t|\epsilon)&=&{1\over\sqrt{2\pi\Delta^2(1-e^{-2\Omega t})}} \no \\
%%&\times&\exp\left[-{(E-\epsilon)^2(1-e^{-\Omega
%%t})^2\over{2\Delta^2(1-e^{-2\Omega t})}}\right].   \label{g2c} \ee
%Then, the solvent diffusion rates can be calculated 
%in Eq.~(\ref{kd1}) by numerical integration. 
%We can rewrite $k_{\rm D}^{i}$ in
%Eq.~(\ref{kd1}) as
%%
%\be (k_{\rm D}^{i})^{-1}=\int_{0}^{\infty}{\rm
%d}t\left[{1\over\sqrt{1-e^{-2\Omega t}}}\exp\left({2 \beta E_{b}^{i} e^{-\Omega
%t}\over{1+e^{-\Omega t}}}\right)-1\right]
%             %&=&\Omega^{-1}f(a),
%              \label{kdint}
%\ee
%
%where $E_{\rm b}^{i}$ is the forward ($i=1$) or backward ($i=2$) reduced
%barrier height, $E_{\rm b}^{i}={(\epsilon\pm \lambda)^2/(4\lambda)}$ with $+$
%sign for $i=1$ and $-$ sign for $i=2$.
% \be f(a)=\int_{0}^{1}{dx\over
%x}\left[{1\over\sqrt{1-x^2}}\exp\left({2ax\over{1+x}}\right)-1\right].
%\label{fa} \ee
%
%Therefore, the diffusional relaxation rate is a product of the bath relaxation
%rate and a function which depends only on the energetics of the electron
%transfer system through the parameter $a$.

In the cases of high and low barrier limits, 
the solvent diffusion rate given in an integral form 
in Eq.~(\ref{kd1}) can be approximately evaluated, 
\be k_{\rm D}^{i}\approx \left\{
\begin{array}{ll} {\Omega\over \ln{2}} \left[1-\left({2\over
\ln{2}}\right){\beta E_{\rm b}^{i}}
%\right.  & \\
%\left.
+\left({\left(2\over\ln{2}\right)}^2-{2\over 3\ln{2}}\right)(\beta E_{\rm b}^i)^2+\cdots\right] & \mbox{$\beta E_{\rm b}^{i}\ll 1$} \\
\Omega \sqrt{\beta E_{\rm b}^{i}\over\pi} \exp(-\beta E_{\rm b}^{i}) &
\mbox{$\beta E_{\rm b}^{i} \gg 1$}
\end{array} \right.
\label{kdapp} \ee
where $E_{\rm b}^{i}$ is the forward ($i=1$) or backward ($i=2$) reduced
barrier height, $E_{\rm b}^{i}={(\epsilon\pm \lambda)^2/(4\lambda)}$.
%with $+$ sign for $i=1$ and $-$ sign for $i=2$.

%It is worthwhile to point out the validity of the Zusman's solution given in
%Ref.~\onlinecite{zusman-cp-80} in terms of nonadiabatic case solution presented
%in this section. 
The Zusman's solution given in
Ref.~\onlinecite{zusman-cp-80} 
amounts to two approximations made in
the nonadiabatic solution. First, it is assumed that electron transfer occurs
only when the solvation polarization energy $E$ matches with the energy
bias $\epsilon$, which yields,
\be K(E)\approx2\pi V^2 \delta(E-\epsilon). \label{kdelta} \ee
In this case, Eq.~(\ref{red1}) is further reduced to
%the integro-differential
%equation for $\rho_{11}$ and $\rho_{22}$ in Eq.~(\ref{dr1dt})
%and (\ref{dr2dt})
%are reduced to
the localized transition model,
%\be
%{\p\over\p t}\rho_{11}(E,t)&=&\cl_{11}\rho_{11}(E,t) \no \\
%                                      &-&2\pi V^2 \delta(E-\epsilon)\{\rho_{11}(E,t)-\rho_{22}(E,t)\}, \label{zus1}\\
%{\p\over\p t}\rho_{22}(E,t)&=&\cl_{22}\rho_{22}(E,t) \no \\
%                                      &+&2\pi V^2 \delta(E-\epsilon)\{\rho_{11}(E,t)-\rho_{22}(E,t)\}, \label{zus2}
%\ee
and the nonadiabatic transition rates $k_{\rm NA}^{i}$ reduces to the original
Marcus expressions given in Eq.~(\ref{kmar}).
%Note that when the solvent relaxation rate is comparable to the thermal energy,
%the solution in Eqs.~(\ref{kf}) and (\ref{kb}) will give the dynamic shift as
%discussed above while Zusman's solution does not.
The second approximation made by Zusman is that the solvent diffusion rates are
approximated by their high-barrier limits given in Eq.~(\ref{kdapp}) for $\beta
E_{\rm b}^{i}\gg1$, which can be written as
\be k_{\rm D}^{i}\approx\Omega|\lambda\pm \epsilon|\rho_{ii}^{\rm
eq}(\epsilon). \label{kd11} \ee

\newpage 

%%%%%%%%%%% REFERENCES %%%%%%%%%%%%%%%%

%\bibliographystyle{../paper/bibtex/aip}
%\bibliography{et}

\begin{thebibliography}{10}

\bibitem{marcus-arpc-64}
R.~A. Marcus,
\newblock Ann.~Rev.~Phys.~Chem. {\bf 15}, 155 (1964).

\bibitem{marcus-bba-85}
R.~A. Marcus and N.~Sutin,
\newblock Biochim. Biophys. Acta. {\bf 811}, 265 (1985).

\bibitem{barbara-jpc-96}
P.~F. Barbara, T.~J. Meyer, and M.~A. Ratner,
\newblock J.~Phys.~Chem. {\bf 100}, 13148 (1996).

\bibitem{vos-nat-93}
M.~H. Vos, F.~Rappaport, J.-C. Lambry, J.~Breton, and J.-L. Martin,
\newblock Nature {\bf 363}, 320 (1993).

\bibitem{jonas-jpc-95}
D.~M. Jonas, S.~E. Bradford, S.~A. Passino, and G.~R. Fleming,
\newblock J. Phys. Chem. {\bf 99}, 2594 (1995).

\bibitem{arnett-jacs-95}
D.~C. Arnett, P.~Vohringer, and N.~F. Scherer,
\newblock J. Am. Chem. Soc. {\bf 117}, 12262 (1995).

\bibitem{kuharski-jcp-88}
{R. A. Kuharski, J. S. Bader, D. Chandler, M. Sprik, M. L. Klein, and R. W.
  Impey},
\newblock J. Chem. Phys. {\bf 89}, 3248 (1988).

\bibitem{bader-jcp-90}
J.~S. Bader, R.~A. Kuharski, and D.~Chandler,
\newblock J. Chem. Phys. {\bf 93}, 230 (1990).

\bibitem{warshel-arpc-91}
A.~Warshel and W.~W. Parson,
\newblock Ann.~Rev.~Phys.~Chem. {\bf 42}, 279 (1991).

\bibitem{zusman-cp-80}
L.~D. Zusman,
\newblock Chem. Phys. {\bf 49}, 295 (1980).

\bibitem{yakobson-cp-80}
B.~I. Yakobson and A.~I. Burshtein,
\newblock Chem.~Phys. {\bf 49}, 385 (1980).

\bibitem{calef-jpc-83}
D.~F. Calef and P.~G. Wolynes,
\newblock J. Phys. Chem. {\bf 87}, 3387 (1983).

\bibitem{garg-jcp-85}
A.~Garg, J.~N. Onuchic, and V.~Ambegaokar,
\newblock J. Chem. Phys. {\bf 83}, 4491 (1985).

\bibitem{hynes-jpc-86}
J.~T. Hynes,
\newblock J. Phys. Chem. {\bf 90}, 3701 (1986).

\bibitem{rips-jcp-87a}
I.~Rips and J.~Jortner,
\newblock J. Chem. Phys. {\bf 87}, 2090 (1987).

\bibitem{sparpaglione-jcp-88}
M.~Sparpaglione and S.~Mukamel,
\newblock J. Chem. Phys. {\bf 88}, 3263 (1988).

\bibitem{yang-jcp-89}
D.~Y. Yang and R.~I. Cukier,
\newblock J. Chem. Phys. {\bf 91}, 281 (1989).

\bibitem{roy-jcp-94}
S.~Roy and B.~Bagchi,
\newblock J.~Chem.~Phys. {\bf 100}, 8802 (1994).

\bibitem{tang-jcp-96}
J.~Tang,
\newblock J.~Chem.~Phys. {\bf 104}, 9408 (1996).

\bibitem{stockburger-jcp-96}
J.~T. Stockburger and C.~H. Mak,
\newblock J.~Chem.~Phys. {\bf 105}, 8126 (1996).

\bibitem{cho-jcp-95}
M.~Cho and R.~J. Silbey,
\newblock J. Chem. Phys. {\bf 103}, 595 (1995).

\bibitem{cao-jcp-00}
J.~Cao and Y.~Jung,
\newblock J. Chem. Phys. {\bf 112}, 4716 (2000).

\bibitem{jortner-acp-99}
J.~Jortner and M.~Bixon, editors,
\newblock {\em Adv.~Chem.~Phys., Electron Transfer from Isolated Molecules to
  Biomolecules}, volume 106 Pts.~1 and 2,
\newblock Wiley, New York, 1999.

\bibitem{gehlen-sci-94}
J.~N. Gehlen, M.~Marchi, and D.~Chandler,
\newblock Science {\bf 263}, 499 (1994).

\bibitem{davis-nat-98}
W.~B. Davis, W.~A. Svec, M.~A. Ratner, and M.~R. Wasielewski,
\newblock Nature {\bf 396}, 60 (1998).

\bibitem{martini-sci-01}
I.~B. Martini, E.~R. Barthel, and B.~Schwartz,
\newblock Science {\bf 293}, 462 (2001).

\bibitem{cukier-arpc-98}
R.~I. Cukier and D.~G. Nocera,
\newblock Ann.~Rev.~Phys.~Chem. {\bf 49}, 337 (1998).

\bibitem{cukier-jpc-96}
R.~I. Cukier,
\newblock J.~Phys.~Chem {\bf 100}, 15428 (1996).

\bibitem{soudackov-jcp-00}
A.~Soudackov and S.~Hammes-Schiffer,
\newblock J.~Chem.~Phys. {\bf 113}, 2385 (2000).

\bibitem{shin-cp-00}
S.~Shin and S.~I. Cho,
\newblock Chem.~Phys. {\bf 259}, 27 (2000).

\bibitem{roberts-jacs-95}
J.~A. Roberts, J.~P. Kirby, and D.~G. Nocera,
\newblock J.~Am.~Chem.~Soc. {\bf 117}, 8051 (1995).

\bibitem{weaver-chemrev-92}
M.~J. Weaver,
\newblock Chem.~Rev. {\bf 92}, 463 (1992).

\bibitem{chen-chemrev-98}
P.~Chen and T.~J. Meyer,
\newblock Chem.~Rev. {\bf 98}, 1439 (1998).

\bibitem{khoshtariya-jpca-01}
D.~E. Khoshtariya, T.~D. Dolidze, L.~D. Zusman, and D.~H. Waldeck,
\newblock J.~Phys.~Chem.~A {\bf 105}, 1818 (2001).

\bibitem{frauenfelder-sci-85}
H.~Frauenfelder and P.~G. Wolynes,
\newblock Science {\bf 229}, 337 (1985).

\bibitem{wolynes-jcp-86}
P.~G. Wolynes,
\newblock J.~Chem.~Phys {\bf 86}, 1957 (1986).

\bibitem{rips-jcp-95}
I.~Rips and E.~Pollak,
\newblock J.~Chem.~Phys. {\bf 103}, 7912 (1995).

\bibitem{rips-jcp-96}
I.~Rips,
\newblock J.~Chem.~Phys. {\bf 104}, 9795 (1996).

\bibitem{alexandrov-znatur-81}
I.~V. Alexandrov,
\newblock Z. Naturforsch A {\bf 36}, 902 (1981).

\bibitem{lucke-jcp-97}
{A. Lucke, C. H. Mak, R. Egger, J. Ankerhold, J. Stockburger, and H. Grabert},
\newblock J. Chem. Phys. {\bf 107}, 8397 (1997).

\bibitem{evans-jcp-98}
D.~G. Evans, A.~Nitzan, and M.~A. Ratner,
\newblock J. Chem. Phys. {\bf 108}, 6387 (1998).

\bibitem{jung-jpca-99}
Y.~Jung, R.~J. Silbey, and J.~Cao,
\newblock J. Phys. Chem. A {\bf 103}, 9460 (1999).

\bibitem{golosov-jcp-01-1}
A.~A. Golosov and D.~R. Reichman,
\newblock J. Chem. Phys. {\bf 115}, 9848 (2001).

\bibitem{golosov-jcp-01-2}
A.~A. Golosov and D.~R. Reichman,
\newblock J. Chem. Phys. {\bf 115}, 9862 (2001).

\bibitem{cao-cpl-99}
J.~Cao,
\newblock Chem. Phys. Lett. {\bf 312}, 606 (1999).

\bibitem{onuchic-jcp-93}
J.~N. Onuchic and P.~G. Wolynes,
\newblock J. Chem. Phys. {\bf 98}, 2218 (1993).

\bibitem{leggett-rmp-87}
{A. J. Leggett, S. Chakravarty, A. T. Dorsey, P. A. Fisher, A. Garg, and W.
  Zwerger},
\newblock Rev. Mod. Phys. {\bf 59}, 1 (1987).

\bibitem{song-jcp-93}
X.~Song and A.~A. Stuchebrukhov,
\newblock J.~Chem.~Phys. {\bf 99}, 969 (1993).

\bibitem{topaler-jcp-94}
M.~Topaler and N.~Makri,
\newblock J.~Chem.~Phys. {\bf 101}, 7500 (1994).

\bibitem{wang-jcp-99}
H.~Wang, X.~Song, D.~Chandler, and W.~H. Miller,
\newblock J.~Chem.~Phys. {\bf 110}, 4828 (1999).

\bibitem{weiss-quant-92}
U.~Weiss,
\newblock {\em Quantum Dissipative Systems},
\newblock World Scientific, Singapore, 1992.

\bibitem{simons-cp-73}
J.~Simons,
\newblock Chem. Phys. {\bf 2}, 27 (1973).

\bibitem{risken-fpe-84}
H.~Risken,
\newblock {\em The Fokker-Planck Equation},
\newblock Springer-Verlag, New York, 1984.

\bibitem{hatano-prb-98}
N.~Hatano and D.~R. Nelson,
\newblock Phys. Rev. B {\bf 53}, 8384 (1998).

\bibitem{dahmen-condmat-99}
K.~A. Dahmen, D.~R. Nelson, and N.~M. Shnerb,
\newblock cond-mat/9903276  (1999).

\bibitem{dahmen-jmb-00}
K.~A. Dahmen, D.~R. Nelson, and N.~M. Shnerb,
\newblock J.~Math.~Biol. {\bf 41}, 1 (2000).

\bibitem{hanggi-rmp-90}
P.~H\"anggi, P.~Talkner, and M.~Borkovec,
\newblock Rev.~Mod.~Phys. {\bf 62}, 251 (1990).

\bibitem{cho-jcp-97}
M.~Cho and R.~J. Silbey,
\newblock J. Chem. Phys. {\bf 106}, 2654 (1997).

\bibitem{makarov-cpl-95}
D.~E. Makarov and M.~Topaler,
\newblock Chem. Phys. Lett. {\bf 245}, 343 (1995).

\bibitem{cao-jcp-95}
J.~Cao, C.~Minichino, and G.~A. Voth,
\newblock J.~Chem.~Phys. {\bf 103}, 1391 (1995).

\bibitem{schwieters-jcp-98}
C.~D. Schwieters and G.~A. Voth,
\newblock J.~Chem.~Phys. {\bf 108}, 1055 (1998).

\bibitem{jang-jcp-01}
S.~Jang and J.~Cao,
\newblock J.~Chem.~Phys. {\bf 114}, 9959 (2001).

\bibitem{casado-jcp-00}
J.~Casado-Pascual, C.~Denk, M.~Morillo, and R.~I. Cukier,
\newblock J.~Chem.~Phys. {\bf 113}, 11176 (2000).

\bibitem{casado-cp-01}
J.~Casado-Pascual, C.~Denk, M.~Morillo, and R.~I. Cukier,
\newblock Chem.~Phys.~ {\bf 268}, 165 (2001).

\bibitem{hartmann-jcp-00}
L.~Hartmann, I.~Goychuk, and P.~H\"anggi,
\newblock J.~Chem.~Phys. {\bf 113}, 11159 (2000).

\bibitem{goychuk-cp-01}
I.~Goychuk, L.~Hartmann, and P.~H\"anggi,
\newblock Chem.~Phys. {\bf 268}, 151 (2001).

\bibitem{mukamel-spec-95}
S.~Mukamel,
\newblock {\em The Principles of Nonlinear Optical Spectroscopy},
\newblock Oxford, London, 1995.

\bibitem{yan-jpc-88}
Y.~Yan, M.~Sparpaglione, and S.~Mukamel,
\newblock J. Phys. Chem. {\bf 92}, 4842 (1988).

\bibitem{reid-jpc-95}
P.~J. Reid, C.~Silva, P.~F. Barbara, L.~Karki, and J.~T. Hupp,
\newblock J. Phys. Chem. {\bf 99}, 2609 (1995).

\bibitem{bader-jcp-94}
J.~S. Bader and B.~J. Berne,
\newblock J.~Chem.~Phys. {\bf 100}, 8359 (1994).

\bibitem{egorov-jpcb-99}
S.~A. Egorov, E.~Rabani, and B.~J. Berne,
\newblock J.~Phys.~Chem.~B {\bf 103}, 10978 (1999).

\bibitem{frantsuzov-cpl-97}
P.~A. Frantsuzov,
\newblock Chem. Phys. Lett. {\bf 267}, 427 (1997).

\bibitem{frantsuzov-jcp-99}
P.~A. Frantsuzov,
\newblock J. Chem. Phys. {\bf 11}, 2075 (1999).

\bibitem{thoss-jcp-01}
M.~Thoss, H.~Wang, and W.~H. Miller,
\newblock  {\bf 115}, 2991 (2001).

\bibitem{jean-jcp-92}
J.~M. Jean, R.~A. Friesner, and G.~R. Fleming,
\newblock J.~Chem.~Phys. {\bf 96}, 5827 (1992).

\bibitem{cao-jcp-97a}
J.~Cao,
\newblock J.~Chem.~Phys. {\bf 107}, 3204 (1997).

\bibitem{tanimura-pra-91}
Y.~Tanimura and P.~G. Wolynes,
\newblock Phys.~Rev.~A {\bf 43}, 4131 (1991).

\bibitem{yan-pra-98}
Y.~J. Yan,
\newblock Phys.~Rev.~A {\bf 58}, 2721 (1998).

\bibitem{caldeira-physa-83}
A.~O. Caldeira and A.~T. Leggett,
\newblock Physica A {\bf 121}, 587 (1983).

\bibitem{rips-jcp-87b}
I.~Rips and J.~Jortner,
\newblock J. Chem. Phys. {\bf 87}, 6513 (1987).

\bibitem{feynmann-statmech-72}
R.~P. Feynmann,
\newblock {\em Statistical Mechanics},
\newblock Addison-Wesley, 1972.

\end{thebibliography}

%\end{multicols}
%\end{document}

\begin{figure}
\caption{ Eigenvalue diagrams are shown for three different values of the
energy bias : (a) $\beta|\epsilon|=0$ (normal regime), 
(b) $\beta|\epsilon|=10$ (activationless regime), and 
(c) $\beta|\epsilon|=20$ (inverted regime) as the coupling constant is increased.
Parameters are $\beta\lambda=10$ and $\beta\Omega=1$. 
%As the coupling constant
%increases, the branching point moves down toward the imaginary axis, and
%two complex branches bifurcate horizontally along the imaginary axis by the
%amount of the Rabi frequency in the strong coupling cases. 
}\label{fig1}
\end{figure}

\begin{figure}
\caption{ Real and imaginary parts of the eigenvalues are shown for the same three
different values as in Fig.~\ref{fig1} : 
(a) $\beta|\epsilon|=0$, (b) $\beta|\epsilon|=10$, and (c) $\beta|\epsilon|=20$. 
Parameters are the same as
those used in Fig.~\ref{fig1}. The real parts of the eigenvalues show
bifurcation and coalescence behaviors as the coupling constant increases, and
the imaginary parts of them asymptotically follow the Rabi oscillation
frequency in the strong coupling limit. }\label{fig2}
\end{figure}

\begin{figure}
\caption{ Comparison of the eigenvalue solution (filled circle)
with other theoretical results;  
the nonadiabatic rate (Eq.~(\ref{kNAtot}), solid line), 
the nonadiabatic-diffusion rate (Eq.~(\ref{ktot}), long dashed line), 
the adiabatic barrier crossing rates calculated from 
the adiabatic eigenvalue (Eq.~(\ref{eigad}), dot-dash line) 
and from the stationary-flux (Eq.~(\ref{flux}), short dashed line)
Solvent relaxation rates
are chosen as (a) $\beta\Omega=0.05$ (slow bath) and (b) $\beta\Omega=1$ (fast
bath), respectively. Other parameters are $\beta\lambda=10$ and
$\beta\epsilon=0$. 
The nonadiabatic 
eigenvalue solution agrees with the nonadiabatic-diffusion rate in the weak coupling case, 
and the agreement with the adiabatic rates at the strong coupling case is qualitative. 
}\label{fig3}
\end{figure}

\begin{figure}
\caption{ Transition between the incoherent and coherent regimes in the
symmetric electron transfer case is manifested in the calculation of the lowest 
eigenvalues. When the coupling constant is about half
of the reorganization energy($\beta\lambda=10$), the higher excited states
cross with the first excited state. Symbols lying on top of the each other
represent the same real parts of the complex conjugate eigenvalues. Other
parameters are $\beta\Omega=1$ and $\beta|\epsilon|=0$. }\label{fig4}
\end{figure}

\begin{figure}
\caption{ 
A phase diagram for the transition between the incoherent and coherent regimes 
in the asymmetric reaction case. Values of $V_c$ where 
two lowest nonzero eigenvalues cross (dot-dash line) are compared with 
an analytical estimate in Eq.~(\ref{trans}) (dashed line). 
Parameters are chosen $\beta \Omega=0.01$ and $\beta \lambda=10$. 
}\label{phase}
\end{figure}

\begin{figure}
\caption{
Eigenvalue solution(filled circle) as well as 
the Marcus rate, Eq.~(\ref{kmartot}),  and the nonadiabatic transition rate, Eq.~(\ref{kNA}), 
weak coupling regime are plotted a a function of the energy bias in 
the slow bath ((a), $\beta\Omega =0.1$) and the fast bath ((b), $\beta\Omega=1$) 
cases. Other parameters are $\beta V=0.01$. 
The dynamical bath effect is evident in the fast bath case in Fig.~\ref{fig5}(b) 
%, while it is not in (c) a fast bath, strong coupling case
%($\beta\Omega=1$,$\beta V=1/2$) in comparison to (d) a slow bath, strong
%coupling case($\beta\Omega=1/9$,$\beta V=1/6$).
% The parameters are chosen such
%that the adiabaticity parameters are the same in the fast and in the slow bath
%cases for the weak coupling ((a) and (b), $\xi=2\pi/100$) and strong coupling
%cases ((c) and (d), $\xi=2\pi/4$), respectively. The reorganization energy is
%$\beta\lambda=10$ in all cases.
}\label{fig5}
\end{figure}

\begin{figure}
\caption{
Dynamical bath effect is estimated by the amount of the peak shift
of the forward nonadiabatic transition rate 
versus the energy bias curve
from its value in the static bath case ($\Omega=0$)
for various values of solvent relaxation rates. 
(a) When the bath is static, the nonadiabatic transition rate
shows a maximum at $|\epsilon|=\lambda$. As the bath relaxation increases,
the maximum position of the nonadiabatic transition rate
shifts into a smaller value of the energy bias. 
(b) The peak shift is plotted as a function of the solvent relaxation rate.
}\label{fig6}
\end{figure}

\begin{figure}
\caption{ Time evolution of populations in the donor and acceptor states are
shown for several values of the coupling constant. The initial wavepacket is
chosen as the equilibrium distribution in the donor state. In weak coupling
cases, (a) and (b), the population relaxes to the equilibrium exponentially,
while in strong coupling cases, (c) and (d), the electronic coherence between
the donor and the acceptor states is observed. }\label{fig7}
\end{figure}

\begin{figure}
\caption{
Breakdown of the positivity in the nonadiabatic diffusion equation
in the case of a large reorganization energy ($\beta\lambda\gg 1$).
A nonzero, real eigenvalue becomes negative 
in the case of a large reorganization energy ($\beta\lambda\ge 12$).
Other parameters are $\beta V=0.1$ and $\beta\Omega=1$.
}\label{fig8}
\end{figure}
%\end{multicols}

%\end{document}

\newpage
%%%%%%%%%% FIGURE SIM %%%%%%%%%%%%%%
\newpage
\begin{figure}
\begin{center}
{\Large{\bf Fig.~1(a) Jung \& Cao}}
\end{center}
\vspace{1in}
\begin{center}
\epsfxsize=\figsize %in
\rotate{\rotate{\rotate{\epsffile{\figdir/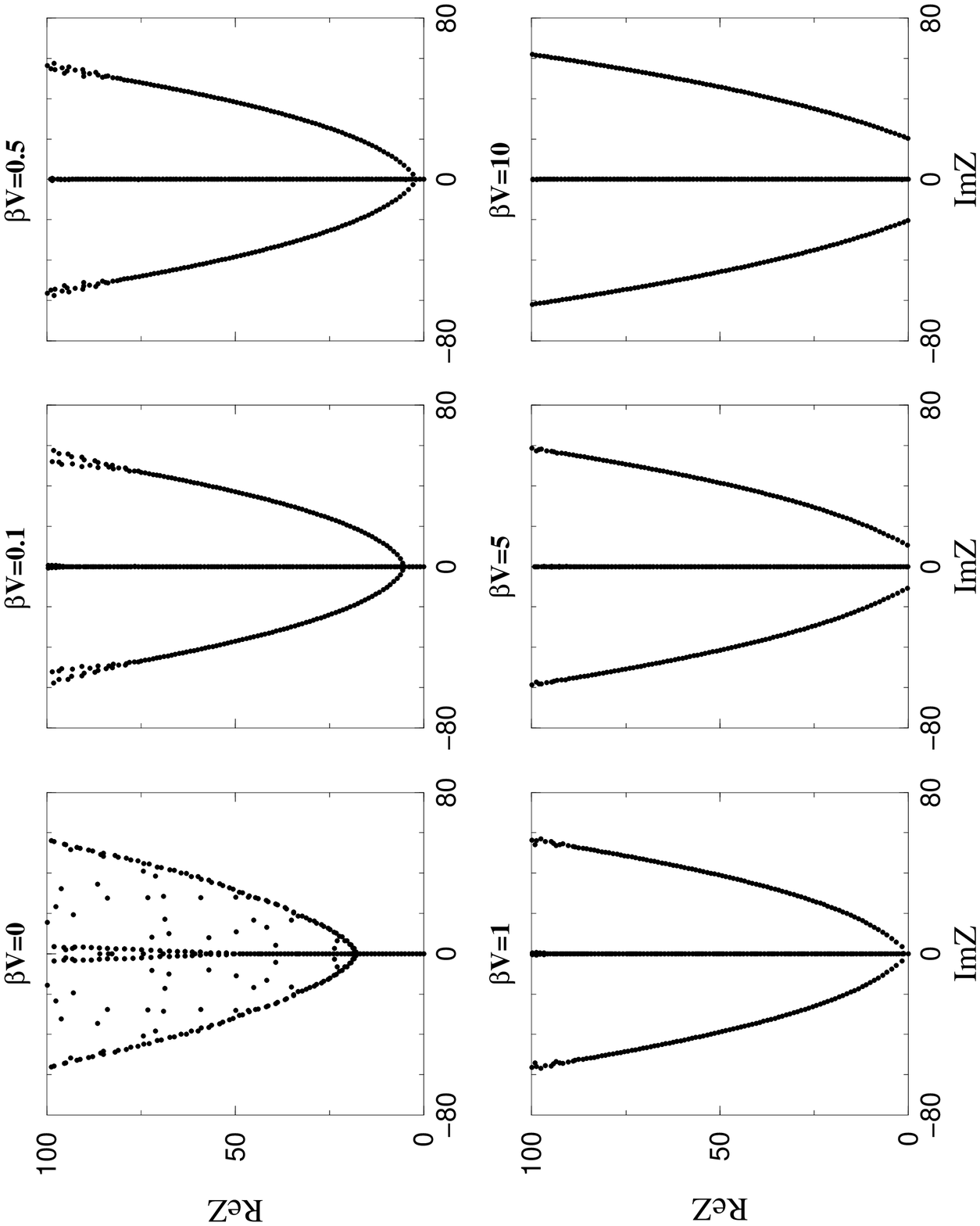}}}}
\end{center}
\end{figure}

\newpage
\begin{figure}
\begin{center}
{\Large{\bf Fig.~1(b) Jung \& Cao}}
\end{center}
\vspace{1in}
\begin{center}
\epsfxsize=\figsize %in
\rotate{\rotate{\rotate{\epsffile{\figdir/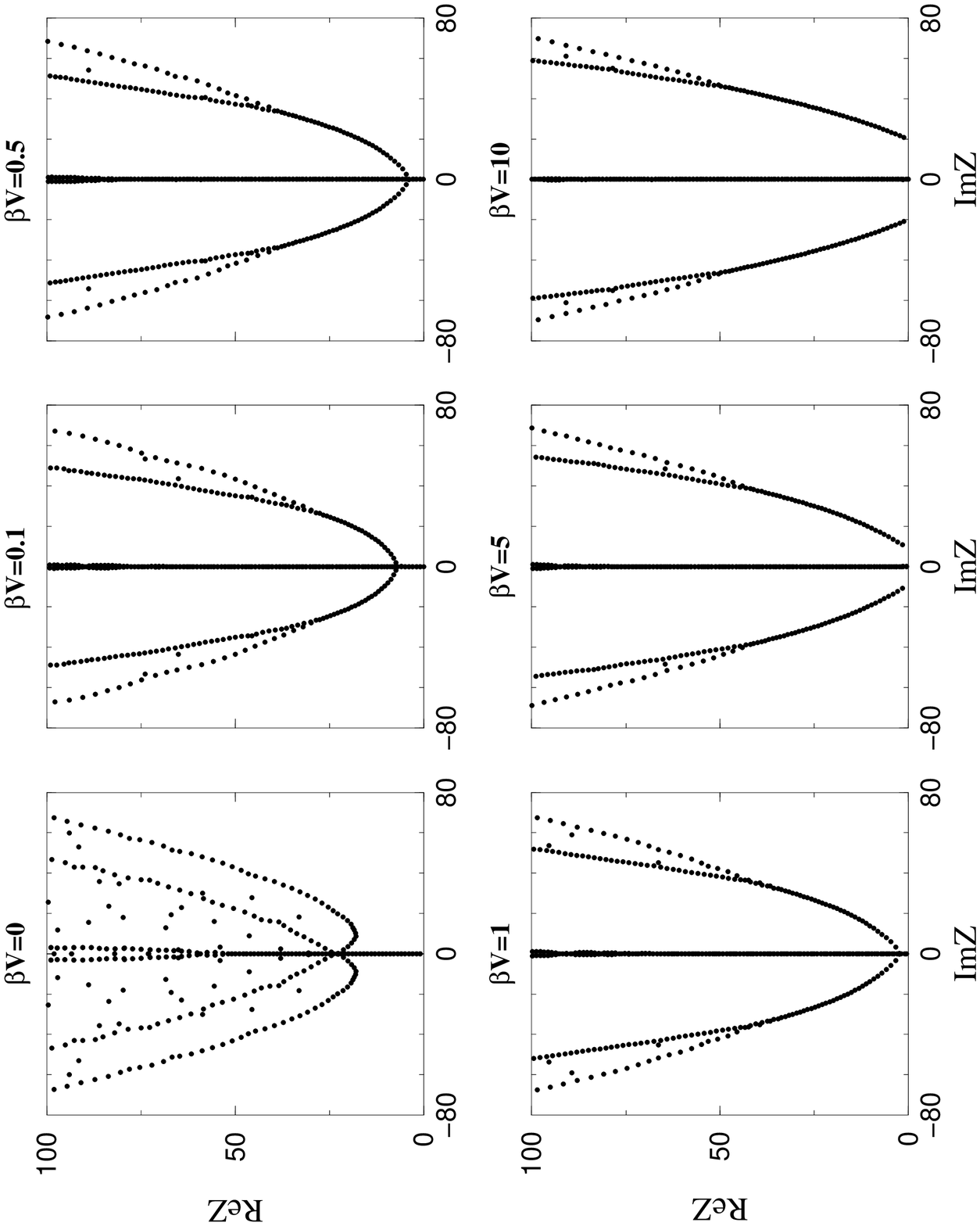}}}}
\end{center}
\end{figure}

\newpage
\begin{figure}
\begin{center}
{\Large{\bf Fig.~1(c) Jung \& Cao}}
\end{center}
\vspace{1in}
\bc \epsfxsize=\figsize %in
\rotate{\rotate{\rotate{\epsffile{\figdir/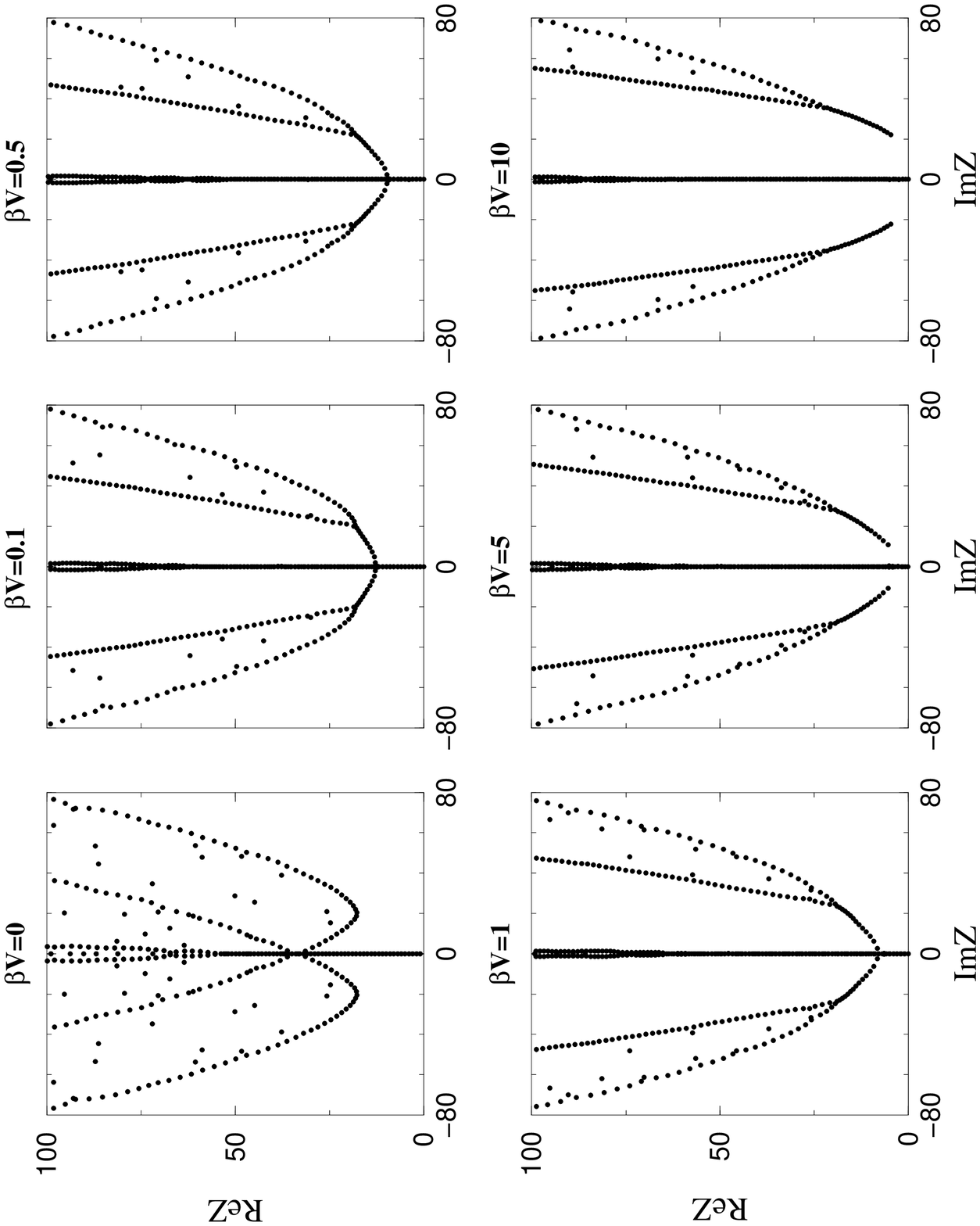}}}} \ec
\end{figure}

\newpage
\begin{figure}
\begin{center}
{\Large{\bf Fig.~2 Jung \& Cao}}
\end{center}
\vspace{1in}
\bc
\epsfxsize=\figsize %in
\rotate{\rotate{\rotate{\epsffile{\figdir/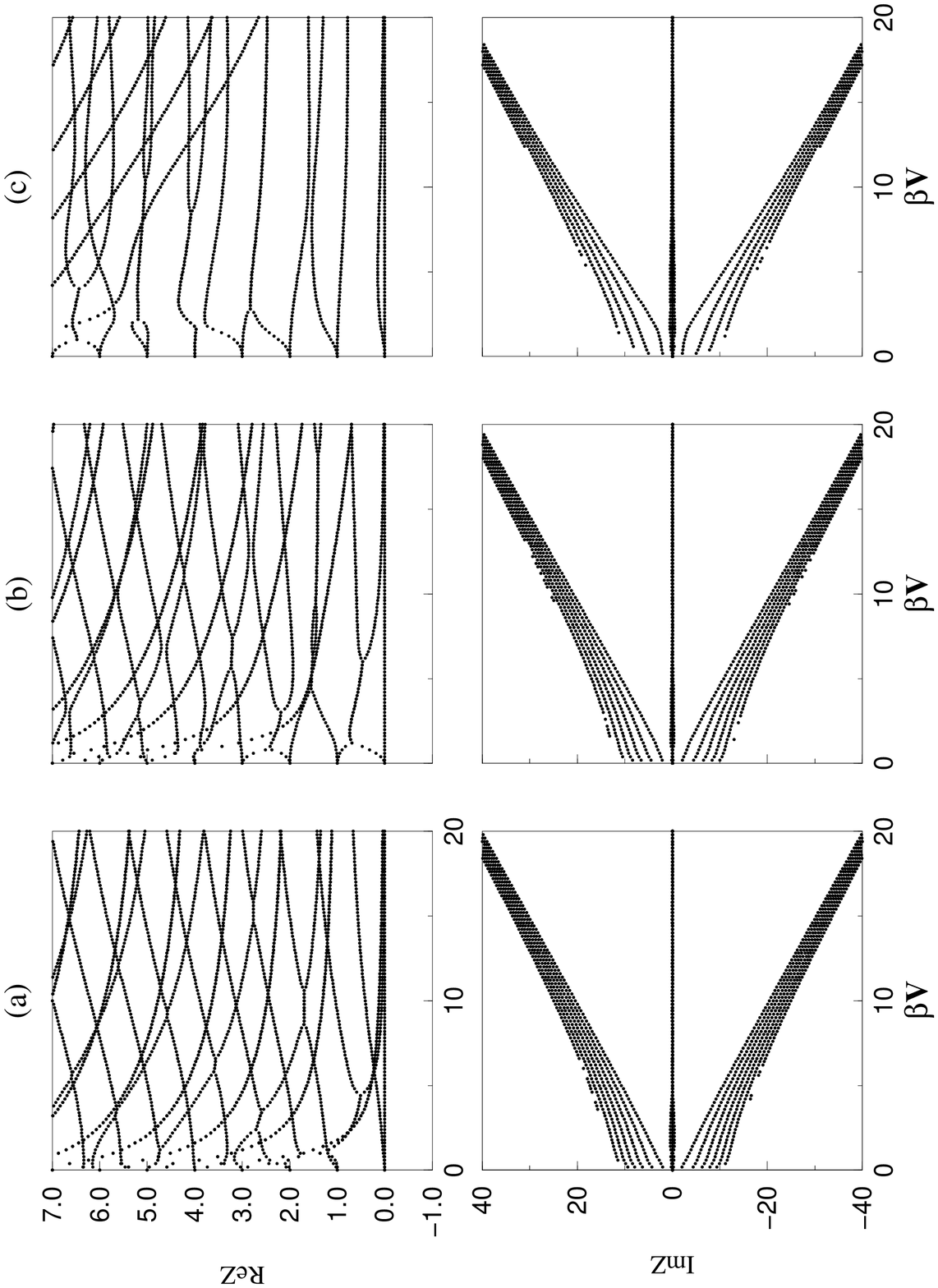}}}}
\ec
\end{figure}

\newpage
\begin{figure}
\begin{center}
{\Large{\bf Fig.~3(a) Jung \& Cao}}
\end{center}
\vspace{1in}
\bc
\epsfxsize=\figsize %in
\epsffile{\figdir/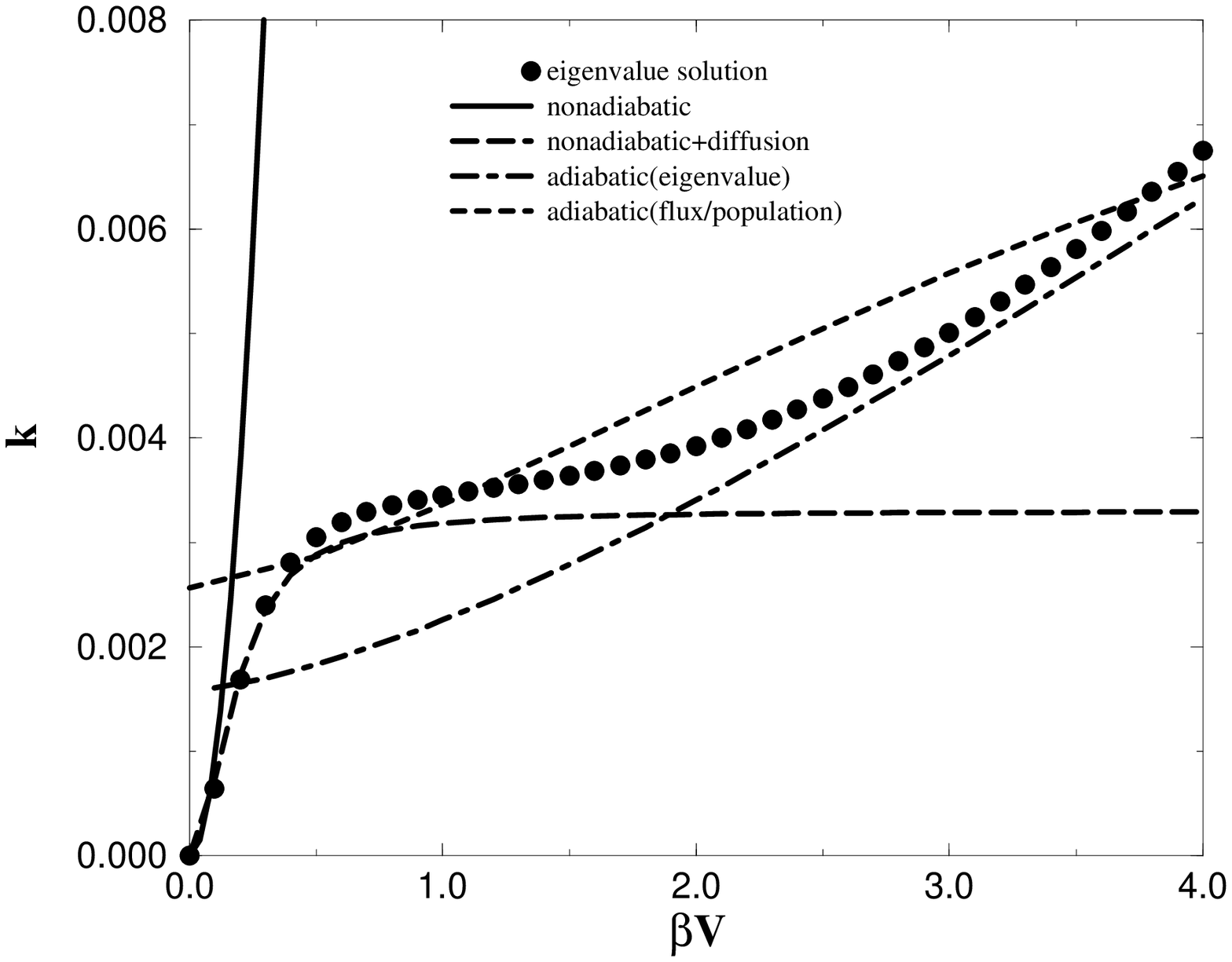}
\ec
\end{figure}

\newpage
\begin{figure}
\begin{center}
{\Large{\bf Fig.~3(b) Jung \& Cao}}
\end{center}
\vspace{1in}
\bc
\epsfxsize=\figsize %in
\epsffile{\figdir/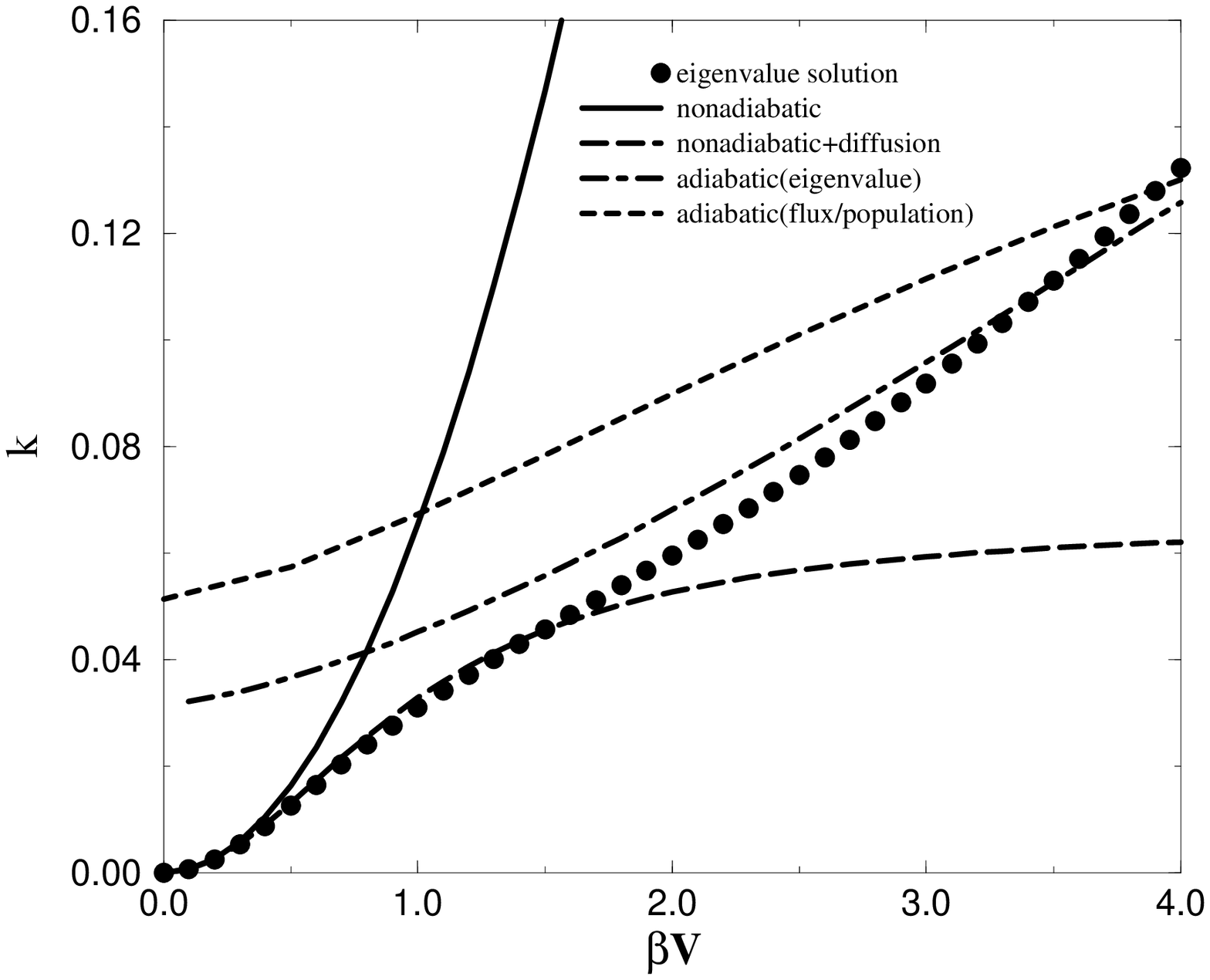}
\ec
\end{figure}

\newpage
\begin{figure}
\begin{center}
{\Large{\bf Fig.~4 Jung \& Cao}}
\end{center}
\vspace{1in}
\bc
\epsfxsize=\figsize %in
\rotate{\rotate{\rotate{\epsffile{\figdir/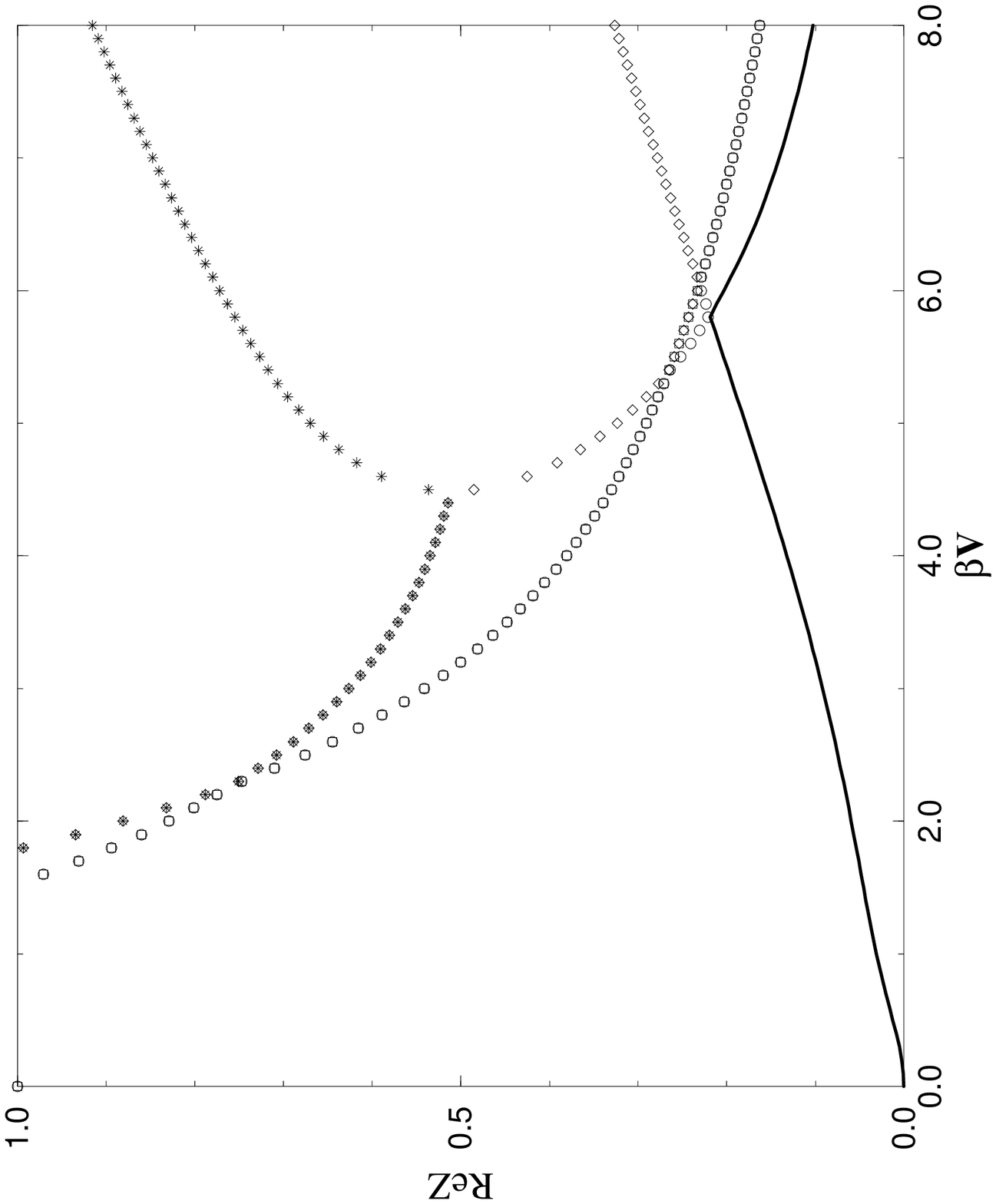}}}}
\ec
\end{figure}

\newpage
\begin{figure}
\begin{center}
{\Large{\bf Fig.~5 Jung \& Cao}}
\end{center}
\vspace{1in}
\bc
\epsfxsize=\figsize %in
\epsffile{\figdir/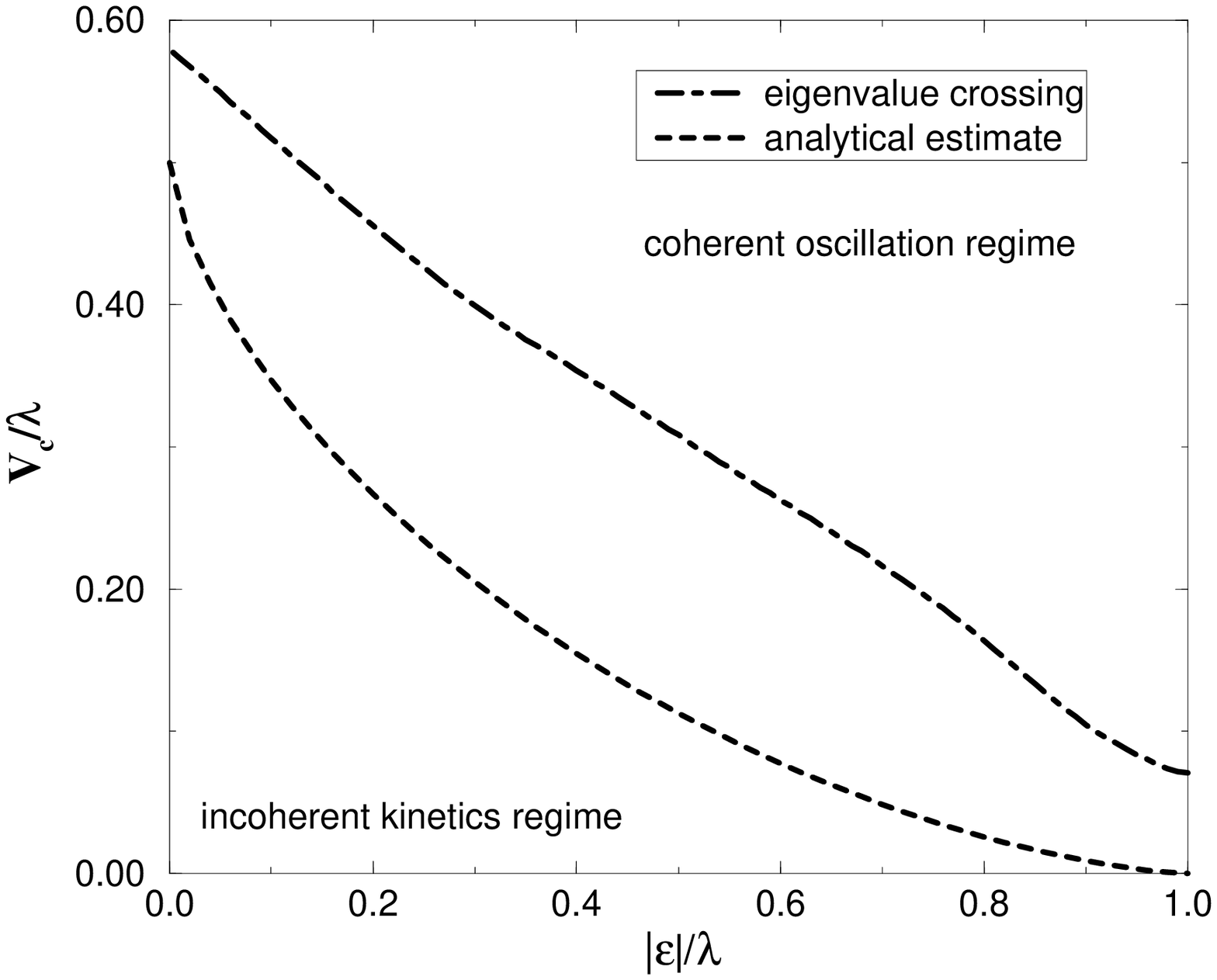}
\ec
\end{figure}

\newpage
\begin{figure}
\begin{center}
{\Large{\bf Fig.~6 Jung \& Cao}}
\end{center}
\vspace{1in}
\epsfxsize=\figsize %in
\epsffile{\figdir/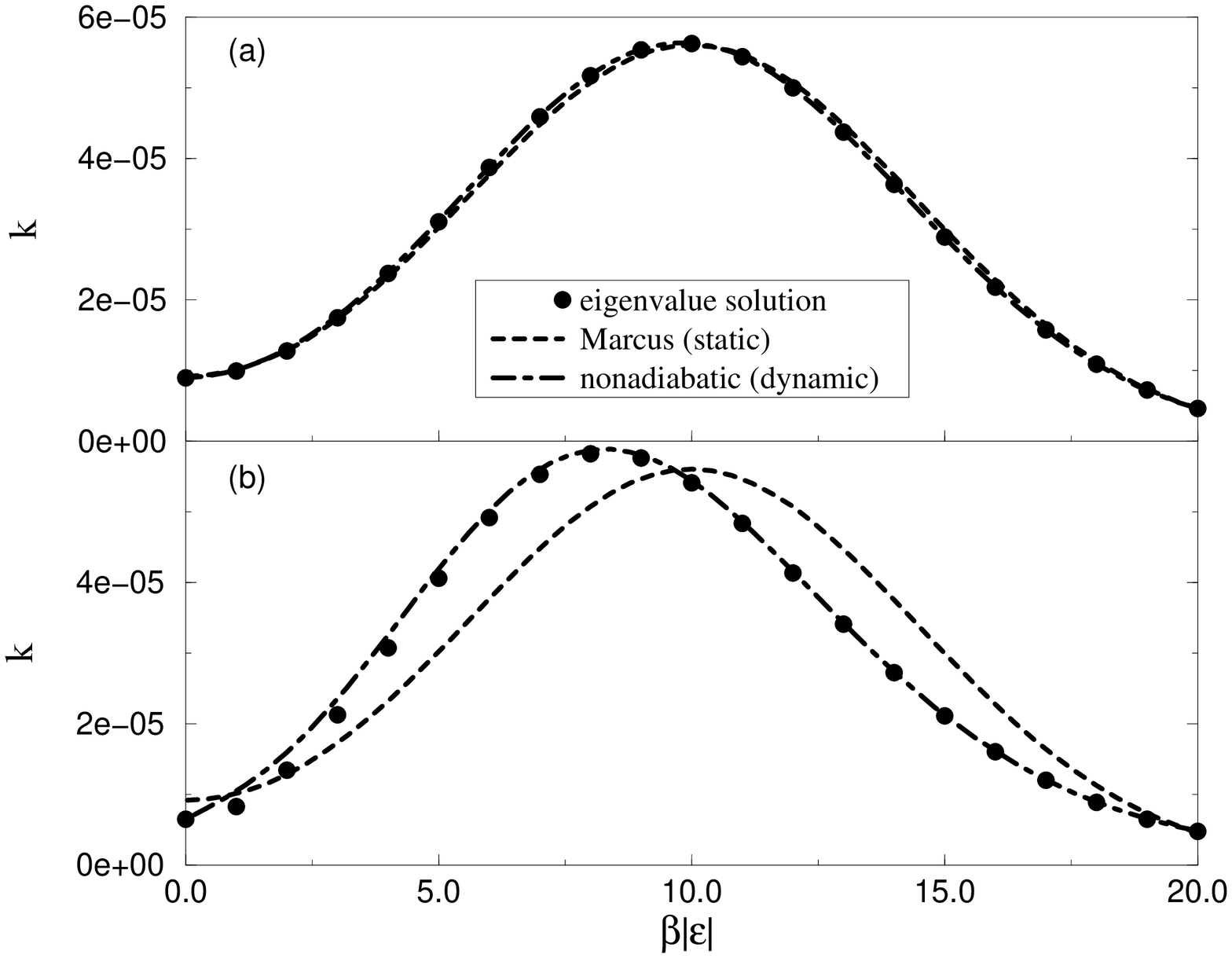}
\end{figure}

%\newpage
%\begin{figure}
%\begin{center}
%{\Large{\bf Fig.~5(b) Jung \& Cao}}
%\end{center}
%\vspace{1in}
%\epsfxsize=5 in
%\rotate{\rotate{\rotate{\epsffile{./fig5b.eps}}}}
%\end{figure}

%\newpage
%\begin{figure}
%\begin{center}
%{\Large{\bf Fig.~5(b) Jung \& Cao}}
%\end{center}
%\vspace{1in}
%\epsfxsize=5 in
%\rotate{\rotate{\rotate{\epsffile{\figdir/fig5b.eps}}}}
%\end{figure}

%\newpage
%\begin{figure}
%\begin{center}
%{\Large{\bf Fig.~5(c) Jung \& Cao}}
%\end{center}
%\vspace{1in}
%\epsfxsize=5 in
%\rotate{\rotate{\rotate{\epsffile{\figdir/fig5c.eps}}}}
%\end{figure}

%\newpage
%\begin{figure}
%\begin{center}
%{\Large{\bf Fig.~5(e) Jung \& Cao}}
%\end{center}
%\vspace{1in}
%\epsfxsize=5 in
%\rotate{\rotate{\rotate{\epsffile{./fig5e.eps}}}}
%\end{figure}

%\newpage
%\begin{figure}
%\begin{center}
%{\Large{\bf Fig.~5(d) Jung \& Cao}}
%\end{center}
%\vspace{1in}
%\epsfxsize=5 in
%\rotate{\rotate{\rotate{\epsffile{\figdir/fig5d.eps}}}}
%\end{figure}

\newpage
\begin{figure}
\begin{center}
{\Large{\bf Fig.~7 Jung \& Cao}}
\end{center}
\vspace{1in}
\bc
\epsfxsize=\figsize %in
{\epsffile{\figdir/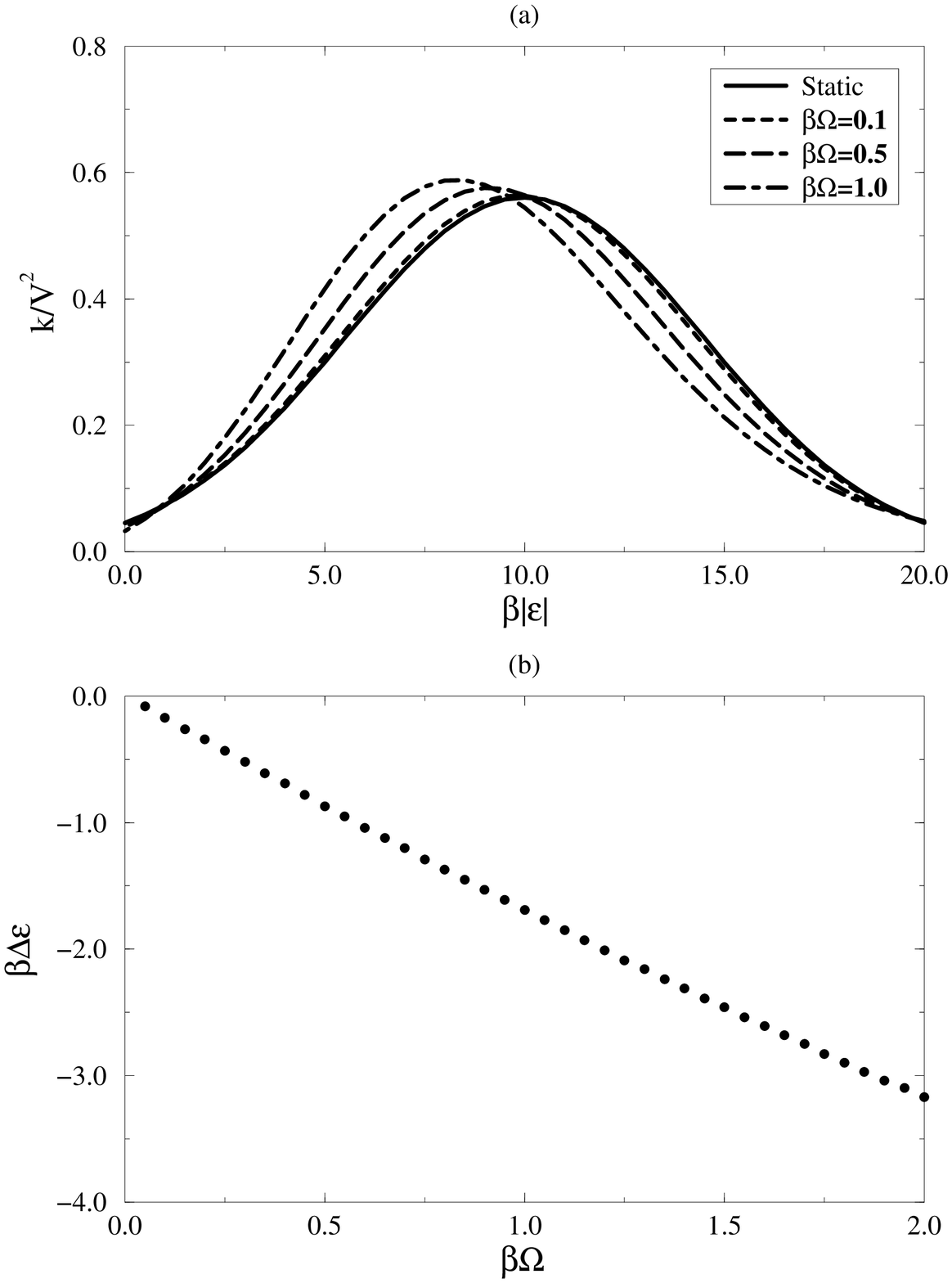}}
\end{center}
\end{figure}

\newpage
\begin{figure}
\begin{center}
{\Large{\bf Fig.~8 Jung \& Cao}}
\end{center}
\vspace{1in}
\bc
\epsfxsize=\figsize %in
\rotate{\rotate{\rotate{\epsffile{\figdir/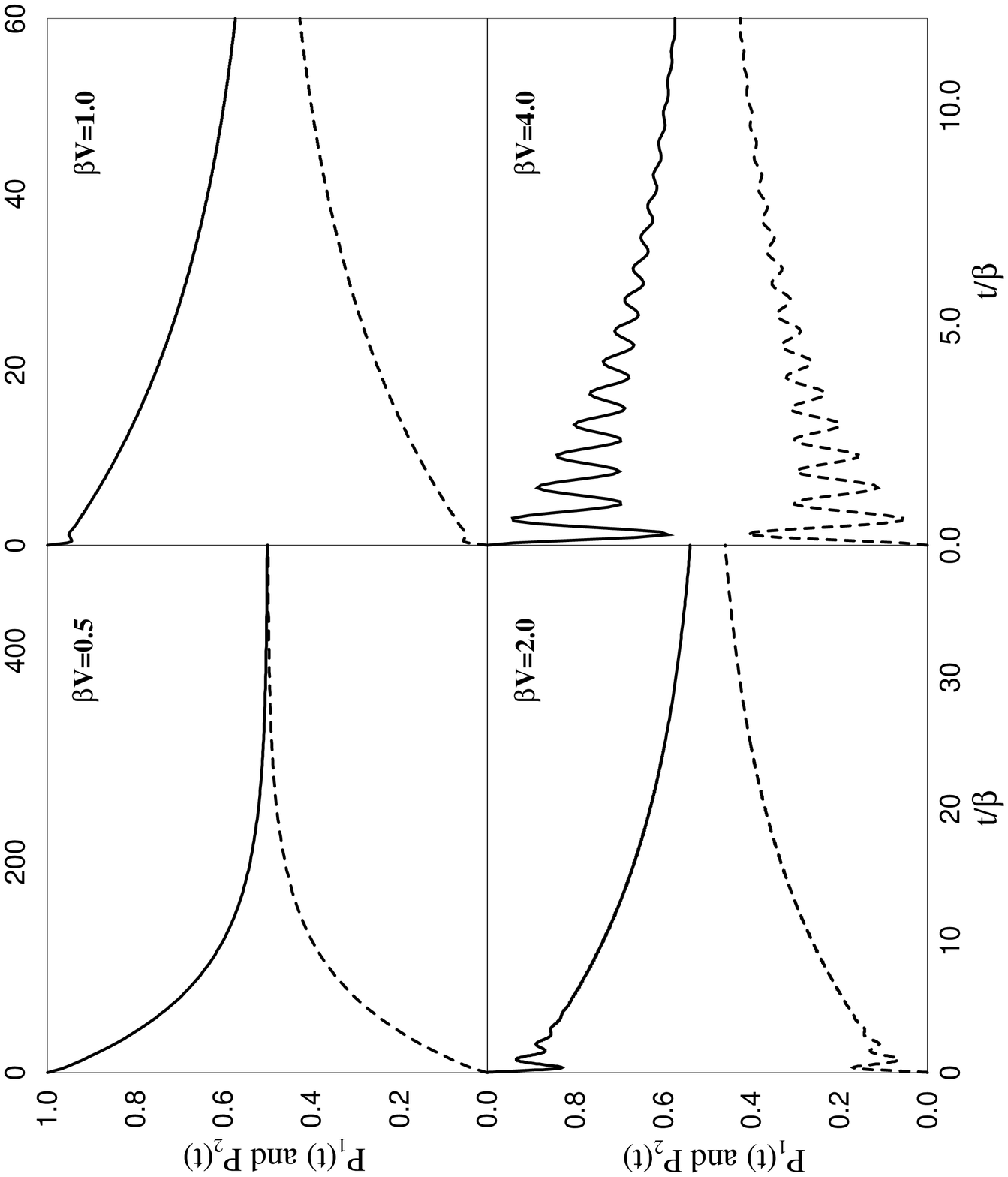}}}}
\ec
\end{figure}

\newpage
\begin{figure}
\begin{center}
{\Large{\bf Fig.~9 Jung \& Cao}}
\end{center}
\vspace{1in}
\begin{center}
\epsfxsize=\figsize %in
\epsffile{\figdir/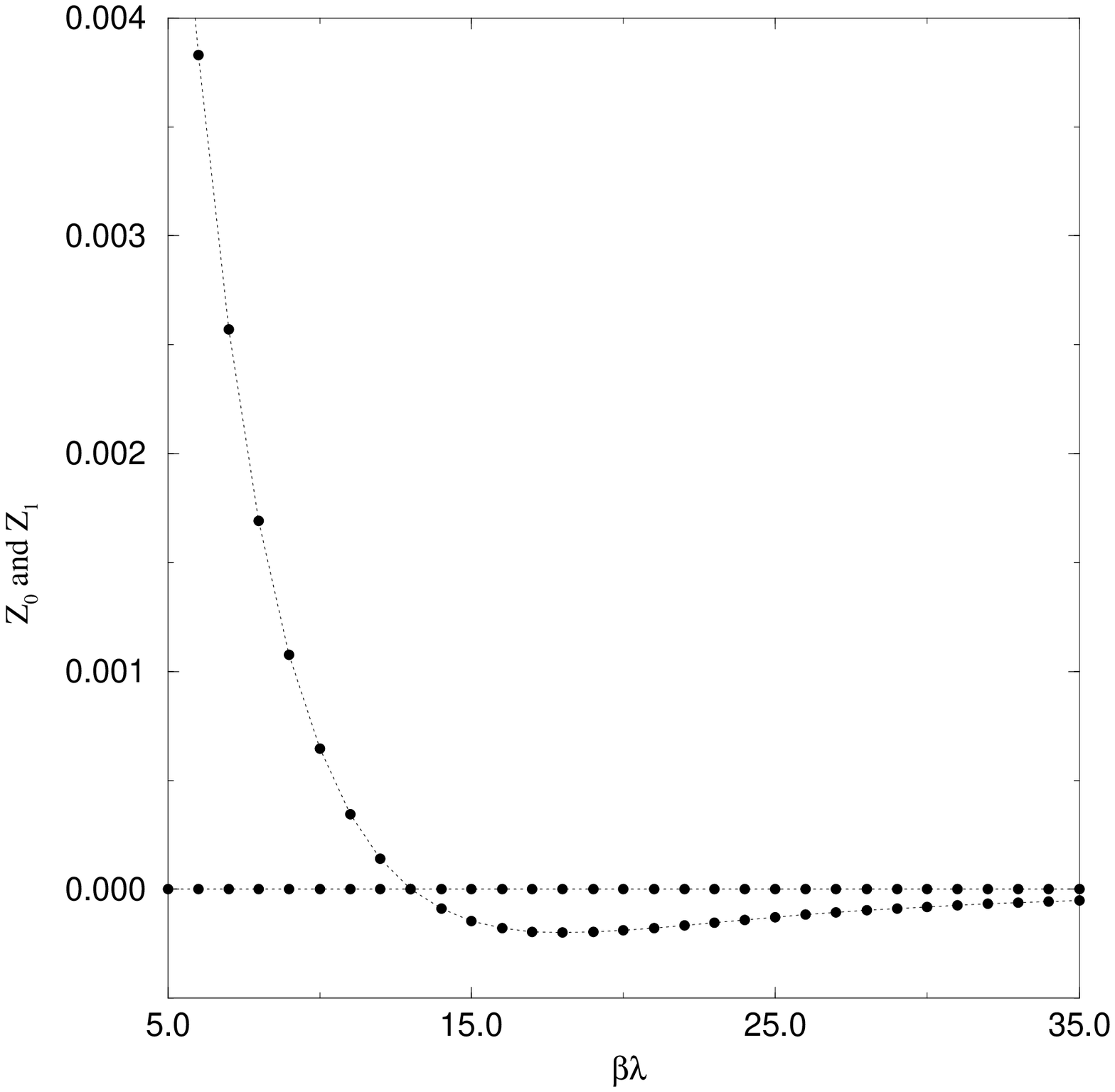}
\end{center}
\end{figure}

\end{document}